\documentstyle[twocolumn,twoside]{ieeetran}
\newcommand\noi{\noindent}
\newcommand\eps{\varepsilon}
\begin{document}

\title{Negative capacitance effect in semiconductor devices}

\author{M.~Ershov, {\it Member}, {\it IEEE},
H.~C.~Liu, L.~Li, M.~Buchanan, Z.~R.~Wasilewski, and A.~K.~Jonscher
\thanks{M.~Ershov is with the
Department of Computer Software, University of Aizu,
Aizu-Wakamatsu City 965-8580, Japan. E-mail: \mbox{ershov@u-aizu.ac.jp} .}
\thanks{H.~C.~Liu, L.~Li, M.~Buchanan, and Z.~R.~Wasilewski are with the
Institute for Microstructural Sciences, National Research Council, Ottawa,
Ontario K1A 0R6, Canada}
\thanks{A.~K.~Jonscher is with the Royal Holloway University of London,
Egham, Surrey, TW20 0EX, UK}
}

\markboth{IEEE Transactions On Electron Devices} %, Vol. XX, No. Y, Month 1999}
{Ershov et al.: Negative capacitance effect in semiconductor devices}

\maketitle

\begin{abstract}
Nontrivial capacitance behavior, including a negative capacitance (NC)
effect, observed in a variety of semiconductor devices, is discussed
emphasizing the physical mechanism and the theoretical interpretation of
experimental data. The correct interpretation of NC can be based on the
analysis of the time-domain transient current in response to a small voltage
step or impulse, involving a self-consistent treatment of all relevant
physical effects (carrier transport, injection, recharging etc.). NC appears
in the case of the non-monotonic or positive-valued behavior of the
time-derivative of the transient current in response to a small voltage step.
The time-domain transient current approach is illustrated by simulation
results and experimental studies of quantum well infrared photodetectors
(QWIPs). The NC effect in QWIPs has been predicted theoretically and
confirmed experimentally. The huge NC phenomenon in QWIPs is due to the
non-equilibrium transient injection from the emitter caused by the properties
of the injection barrier and the inertia of the QW recharging.
\end{abstract}

%\begin{keywords}
%\end{keywords}

\section{Introduction}
Capacitance characteristics provide a powerful spectroscopic method for the
non-destructive testing of semiconductor devices and evaluation of their
structural and physical parameters. Capacitance or admittance spectroscopy
also give an insight into the device physics, provided that the experimental
data are correctly interpreted. Quite often the capacitance exhibits highly
non-trivial characteristics, most notable of which is the phenomenon of
negative capacitance (NC). The NC effect has been displayed by a variety of
electronic devices, both heterostructures and homostructures, made of
crystalline or amorphous semiconductors, such as Si, Ge, GaAs, HgCdTe, Se,
and
others~\cite{KanaiNC55,MisawaNC57,MisawaNR66a,MisawaNR66b,KazarinovNC67,MarshNC67,VogelNC69,RockstadNC71,AllisonNC71,AltunyanNC71,BrodovoiNC73,KolomietsNC74,EgiazaryanNC75,DeshevoiNC77,VeingerNC78,NoguchiNC80,NadkarniNC83,AlimpievNC84,BlatterNC86,JonscherNC86,FuNC87,JonscherNC88,WernerNC88,WuNC89,HeNC89,WuNC90,ChampnessNC90,MerlinNC90,MuretNC91,MorantNC92,BealeNC92a,BealeNC92b,BoltaevNC95,ButtPRL96,KuttyNC96,ErshovCapAPL,MisiakosNC97,OmuraNC97}.
These devices include p-n junctions, Schottky diodes, metal-insulator-metal
devices, MESFETs, metal-insulator-semiconductor structures, weakly coupled
superlattices, quantum mesoscopic devices, quantum well infrared
photodetectors, and so on. Microscopic physical mechanisms of NC in different
devices are, obviously, different, but there should be some general principle
behind NC common to all types of devices. Although the physical origin of NC
has been discussed in the literature~\cite{Jonscher86,JonscherURL}, the
concept of NC is still not widely recognized. Moreover, there is no adequate
discussion of capacitance in classical texts on physics of semiconductor
devices. The NC effect reported in the literature has been often referred to
as ``anomalous'' or ``abnormal''. NC measured experimentally has been
sometimes (incorrectly) attributed to instrumental problems, such as
parasitic inductance~\cite{Butcher96} or poor measurement equipment
calibration~\cite{Huang97}. Regrettably, in many cases experimental NC data
were not reported in the literature due to the confusion caused by the NC
effect. On the other hand, theoretical interpretations of the NC phenomenon
were often based on considerations of purely electrostatic charge
redistribution inside the device. However, a simple incremental charge method
of capacitance calculation can be incorrect in the case of large conduction
current (which often accompanies NC phenomenon), and more rigorous approaches
considering transient response in the time or frequency domain should be
used.

Recently, the NC effect in homogeneous (barrier-free) semiconductor
structures has been considered theoretically in detail in
ref.~\cite{Negcap96}. It was shown that NC can appear if the conductivity is
inertial (i.~e. current lags behind voltage) and the reactive component of
the conduction current is larger than the displacement current. This
situation can occur, for example, in structures with Drude conductivity, or
in the case of impact ionization of impurity atoms. The real devices,
however, contain contacts, which can influence strongly the small-signal
characteristics and result in NC. Indeed, it was verified that in many cases
the NC phenomenon was caused by the contact or interface effects (see, for
example, ref.~\cite{JonscherNC88}).

In this work, we discuss the NC effect with an emphasis on the theoretical
interpretation of this phenomenon. We obtain the general relationships
between the transient current in the time-domain and capacitance in the
frequency domain. These relationships, following from the properties of the
Fourier transform, are independent of the particular physical processes and
applicable to all types of electronic devices. The origin of NC is related to
the non-monotonic or positive-valued behavior of the time-derivative of
transient current upon application of a voltage step. We present the results
of simulation and experimental studies of quantum well infrared
photodetectors (QWIPs) displaying a huge NC. In QWIPs, this effect is due to
the non-equilibrium transient injection from the emitter, caused by the
properties of the injecting contact and inertia of the QW's recharging
processes.

\section{Definition of capacitance and methods of its calculation}
For simplicity, we consider two-terminal semiconductor devices.
Capacitance is defined as
\begin{equation}
C(\omega)={1\over\omega} \rm{Im} \left[ Y(\omega) \right],
\label{cap-def}
\end{equation}

\noi where
\begin{equation}
Y(\omega)=\frac{\delta I(\omega)}{\delta V(\omega)}
\label{imp-def}
\end{equation}

\noi is the device admittance relating the small-signal harmonic current
flowing through the terminals and small-signal voltage ($\delta I$, $\delta V
\sim e^{i\omega t}$). The real part of the complex admittance is called
conductance $G(\omega)=\rm{Re} \left[ Y(\omega) \right]$. In general,
capacitance calculations should involve the solution of the system of
equations describing device operation (Poisson equation, current continuity
equation, etc.) in the time or frequency domain. There are several
established methods for calculation of the capacitance (and, more generally,
of the admittance)~\cite{LauxSS85}.

\subsection{Incremental charge approach}
In the incremental charge approach, steady-state equations describing device
operation are solved for voltages $V$ and $V+\Delta V$ ($\Delta V$ is small).
The incremental charge distribution in the device $\delta Q(x)$ is separated
into positive and negative components $\Delta Q$ and $-\Delta Q$, which are
assigned to the respective contacts (the net charge increment inside the
device is zero, according to the Gauss theorem). The capacitance is then
calculated as $C=\Delta Q /\Delta V$. The main advantage of this method is
its simplicity, as it requires only a steady-state simulation program for the
calculation. However, there are several difficulties with this approach. One
of the problems is that there is no rigorous procedure for the separation of
incremental charge distribution into positive and negative parts, and its
assignment to the contacts. This problem becomes especially important in the
case of devices with more than two contacts and two- or three-dimensional
geometry, as well as in the case when the incremental charge density of
electrons and holes is distributed across the device area. Several heuristic
approaches to tackle this problem have been proposed, without rigorous
justification. Furthermore, this method allows the calculation only of the
low-frequency value of the capacitance. More importantly, the incremental
charge approach can be rigorously justified only if there is no conduction
current in the device under DC conditions, which will be shown below. This
approach can work well in the case of very low conduction current (as in
reverse biased p-n junctions or Schottky diodes, MOS structures etc.). In the
general case of non-zero DC current in the device, the incremental charge
approach can fail, and its applicability in each particular case should be
carefully examined.

\subsection{Sinusoidal steady-state analysis}
In the sinusoidal steady-state analysis (SSSA) approach, the system of the
time-dependent equations is linearized around a steady-state solution for
harmonic small-signal quantities, and then solved for a particular frequency.
The SSSA method is rigorous, rather simple to carry out, and fast. The
disadvantage is that it requires solution of a system of equations for each
frequency to obtain the frequency dependence of capacitance.

\subsection{Method based on Fourier analysis}
The method based on the Fourier analysis involves calculation of the
transient response of the device to a small time-dependent voltage excitation
(usually in the form of a step-function) applied at time $t=0$. Admittance is
calculated as the ratio of the Fourier components of the transient current
$\delta I(t)=I(t)-I(0^-)$ and voltage $\delta V(t)=V(t)-V(0^-)$. The
amplitude of the transient voltage should be chosen small enough to ensure
the linearity of the transient effects. The particular value of the amplitude
is dictated by the problem under consideration (it can be much less than the
total applied voltage, thermal voltage, etc.) On the other hand, it should
not be large enough that the transient current is properly resolved. This
method is rigorous. It requires the solution of the transient problem only
once to calculate the capacitance and conductance for any frequency. Its
disadvantage is the stringent requirement on the choice of the time steps to
obtain the proper frequency resolution and numerical
accuracy~\cite{LauxSS85}. The Fourier analysis method translates the
small-signal problem from the frequency domain into the time domain. Although
both the frequency-domain and time-domain representations are appropriate,
the time-domain approach sometimes allows a more clear interpretation of the
observed results in terms of the relevant physical effects. This method will
be used in the next sections to relate the properties of the
capacitance-frequency (C-F) characteristics to the time-domain behavior of
the transient current, and to explain the origin of NC.

\section{The origin of negative capacitance}
Let us consider transient current in a semiconductor device in response to
an applied voltage step (see Fig.~1):
\begin{equation}
\delta V(t) \equiv V(t) -
V(0^-) = \Delta V  \theta(t),
\label{trV}
\end{equation}

\begin{eqnarray}
\label{trI}
\delta I(t) & \equiv & I(t) - I(0^-) = \nonumber \\\
&&[I(t)-I(\infty)]\,\theta(t)+
[I(\infty)-I(0^-)] \, \theta(t),
\end{eqnarray}

\noi
where $\theta(t)$ is the unity step function.  The quantities with ``$+$''
and ``$-$'' superscripts denote single-sided values of the discontinuous
functions, for example, $V(0^-)=\lim_{\tau\rightarrow 0} V(\tau),
\tau<0$. In equation (\ref{trI}) we decomposed the transient current $\delta
I(t)$ into the step-like component (DC conductivity) and transient current
$\delta J(t) = [I(t)-I(\infty)]\,\theta(t)$, so that $\delta J(t)\rightarrow
0$ as $t \rightarrow \infty$. Substituting the Fourier expansions of
eqs.~(\ref{trV}) and (\ref{trI}) into formula~(\ref{imp-def}), and noting
that $\int_{-\infty}^{\infty} \theta(t) e^{-i\omega t} dt = 1/(i\omega)$
(strictly speaking, to ensure convergence of this integral, we must add to
$\omega$ an infinitesimal negative imaginary part, i.~e. replace $\omega$ by
$\omega-i0$), we obtain the following expression for admittance:

\begin{equation}\label{adm-step}
Y(\omega)=i\omega \int_{0}^{\infty} \delta I(t)\, e^{-i\omega t}\, dt.
\end{equation}

\noi Separating the real and imaginary parts in formula~(\ref{adm-step}), we
obtain the expressions for the capacitance and conductance:

\begin{equation}
C(\omega)={1 \over {\Delta V}} \int_0^{\infty} \delta J(t)
\cos\omega t \,dt.  \label{cap-eq}
\end{equation}

\begin{equation}
G(\omega)=\frac{I(\infty)-I(0^-)}{\Delta V} +{\omega \over {\Delta V}}
\int_0^{\infty} \delta J(t) \sin \omega t \,dt .
\label{cond-eq}
\end{equation}

In general, the transient current $\delta J(t)$ contains an impulse-like
component and a slowly varying relaxation component (Fig.~1):
\begin{equation} \delta J(t)=C_0 \Delta V \delta(t) + \delta j(t).
\label{delJ} \end{equation}

\noi Here $\delta(t)$ is the delta-function. The impulse-like component
corresponds to a current charging the geometric capacitance $C_0$ (which is
also called a feedthrough capacitance or ``cold'' capacitance), assuming that
application of a voltage step results in an instantaneous change of charges
on the contacts. Physically, this current is due to the displacement current
in the semiconductor. For a device with a 1D geometry, we have
$C_0=\eps\eps_0A/L$, where $\eps$ is the (average) relative dielectric
permittivity, $\eps_0$ is the permittivity of vacuum, $A$ is the area, and
$L$ is the distance between the contacts. The relaxation component $\delta
j(t)$ can be due to the electron transport, trapping, impact ionization, and
other physical processes. Substituting formula~(\ref{delJ}) into
eqs.~(\ref{cap-eq}) and (\ref{cond-eq}), we obtain the equations for
capacitance and conductance in terms of the transient relaxation current
$\delta j(t)$:
\begin{equation}
C(\omega)=C_0 +{1\over{\Delta V}} \int_0^{\infty} \delta j(t)\, \cos\omega t \,
dt .
\label{cap-EQ}
\end{equation}

\begin{equation}
G(\omega)=G(0)+\frac{\omega}{\Delta V} \int_0^{\infty} \delta j(t)\, \sin\omega
t\, dt,
\label{cond-EQ}
\end{equation}

\noi where $G(0)=(I(\infty)-I(0^-)/\Delta V$ is the DC or steady-state
conductivity.
It is useful to obtain an alternative formulation of eqs.~(\ref{cap-EQ}) and
(\ref{cond-EQ}). Using integration by parts, we get:
\begin{equation}
C(\omega)=C_0 +{1\over{\omega\Delta V}} \int_0^{\infty}
\left[-\frac{d\delta j(t)}{dt} \right] \, \sin\omega t \, dt .
\label{cap-EQA}
\end{equation}

\begin{equation}
G(\omega)=G(\infty)+\frac{1}{\Delta V} \int_0^{\infty} \left[\frac{d\delta
j(t)}{dt} \right] \, \cos
\omega t
\, dt,
\label{cond-EQA}
\end{equation}

\noi where $G(\infty)=[I(0^+)-I(0^-)]/\Delta V$ is the high-frequency
conductivity. It can be shown that the derivative of the relaxation current
$\delta h(t)=d\delta j(t)/dt$ corresponds to the relaxation component of
transient current in response to a voltage impulse (response function).
Indeed, if $\delta H(t)$ is the transient current in response to a voltage
impulse $\delta V(t)=v\,\delta(t)$ ($v$ is the ``power'' of the voltage
impulse), then the admittance is given by the formula
$Y(\omega)=1/v\int_0^{\infty}\delta H(t) e^{-i\omega t} dt$. Comparing this
expression with formula~(\ref{adm-step}), we obtain:
\begin{equation}\label{tr-cur-imp}
\delta H(t)=\frac{v}{\Delta V}
\left\{ C_0\Delta V
\frac{d\delta(t)}{dt}
+[I(0^+)-I(0^-)]\,\delta(t) +\frac{d\delta j(t)}{dt}\right\}.
\end{equation}

\noi The first term in eq.~(\ref{tr-cur-imp}) corresponds to the displacement
current, the second term corresponds to the instantaneous current response due
to the high-frequency conductivity, and the last term is related to the
relaxation current (see Fig.~2).

It is seen from eqs.~(\ref{cap-EQ})--(\ref{cond-EQA}) that the frequency
dependence of capacitance and conductance is determined by the cosine and
sine transforms of the transient current $\delta j(t)$ (and $\delta h(t)$),
and, therefore, by the time-domain behavior of the transient current. If the
function $\delta j(t)$ is positive-valued, and decreases monotonically and
smoothly (without inflections) to zero as $t\rightarrow \infty$, then the
integral in eq.~(\ref{cap-EQ}) is positive, and the capacitance $C(\omega)$
is larger than $C_0$ at any frequency. Indeed, in this case the function
$-d\delta j(t)/dt$ is positive and monotonically decreases to zero, so that
the positive contribution to the integral in eq.~(\ref{cap-EQA}) over the
first half of a sine period overweighs the negative contribution over the
second half of a period (see Fig.~3). On the other hand, if the function
$-d\delta j(t)/dt$ is negative and monotonically increasing to zero, then the
integral in eq.~(\ref{cap-EQA}) is negative, capacitance $C(\omega)$ is less
than $C_0$, and can be negative. In the case of non-monotonic or
positive-valued behavior of the derivative of the transient current, the
capacitance can be negative in limited frequency ranges.

Thus, {\em the origin of NC is related to the non-monotonic or
positive-valued behavior of the derivative of the transient current in
response to a small voltage step}, as was first proposed by
Jonscher~\cite{Jonscher86}.

\section{Relations between the transient current and admittance}

Let us consider some other important relations between the frequency-dependent
admittance and transient current following from the mathematical properties of
the Fourier transform.

The low-frequency capacitance value is given by the following formula:
\begin{equation}
C(0)=C_0+\frac{1}{\Delta V} \int_0^{\infty} \delta j(t)\, dt.
\label{cap-lf}
\end{equation}

\noi
Thus, the value of the low-frequency incremental capacitance $\Delta
C(0)=C(0)-C_0$ is determined by the net area under the curve $\delta j(t)$. If
the net area is negative, the low-frequency capacitance $C(0)$ is less than
$C_0$ and can be negative.

A sum rule complementary to eq.~(\ref{cap-lf}) reads:
\begin{equation}
\delta j(0^+) = \frac{2}{\pi} \int_0^{\infty} [C(\omega)-C_0]\, d\omega,
\label{sum}
\end{equation}

\noi which follows immediately from the following relation for a function
$f(t)$ such that $f(t)=0$ if $t<0$:
\begin{equation}
\int_0^{\infty} \rm{Re}[f(\omega)] \,d\omega = \sqrt{\frac{\pi}{8}}\, f(0^+).
\label{sum-rel}
\end{equation}

\noi Therefore, the value of the relaxation current $\delta j$ at $t=0^+$
determines the total area under the curve of incremental capacitance $\Delta
C(\omega) = C(\omega)-C_0$.

The high-frequency capacitance value is equal to the geometric capacitance:
\begin{equation}
C(\infty)=C_0,
\label{cap-hf}
\end{equation}

\noi because the integral of the product of a slowly varying function and
fast oscillating harmonic function tends to zero as $\omega
\rightarrow\infty$. From the physical viewpoint, this is due to the fact that
at high frequencies the physical processes related to electron transport are
``frozen'' due to the finite inertia, and the total small-signal current
at high frequencies contains only the displacement component associated with the charging of the
geometric capacitance.

Let us further consider the asymptotic behavior of capacitance at high
frequencies. Integrating eq.~(\ref{cap-EQ}) by parts, we obtain:

\begin{eqnarray}
C(\omega)&=&C_0+\frac{1}{\Delta V}\, \delta j(t)\, \frac{\sin(\omega t)}{\omega}
\bigg|_0^{\infty}
\nonumber \\\
&& -\frac{1}{\omega\Delta V} \int_0^{\infty} t\, \frac{d \delta
j(t)}{dt} \, \frac{\sin\omega t}{t}\, dt.  \label{parts}
\end{eqnarray}

\noi Since $\delta j(\infty)=0$, the second term in (\ref{parts})
disappears. Further, by utilizing the following theorem for the function
$f(t)$ satisfying Dirichlet's conditions ($f(t)$ has a finite number of
maxima and minima in the interval, and has only a finite number of finite
discontinuities):

\begin{equation}
\lim_{\omega \rightarrow \infty} \int_0^{\infty} f(t)\,
\frac{\sin(\omega t)}{t} \, dt = \frac{\pi}{2} f(0^+) ,
\label{theorem}
\end{equation}

\noi we obtain the following relationship:
\begin{equation}
\lim_{ \omega \rightarrow \infty } \left\{ \omega [C(\omega)-C_0] \right\} =
-\frac{1}{\Delta V} \frac{\pi}{2} \left( t \frac{d \delta j} {d t}
\right)_{t=0^+} .
\label{rel}
\end{equation}

\noi Therefore, the asymptotic behavior of the capacitance at $\omega
\rightarrow \infty$ is determined by the sign of the time derivative of the
relaxation current at $t=0^+$. $C(\omega)$ approaches $C_0$ as $\omega
\rightarrow \infty$ from above  if $\left( d \delta j/dt\right)
_{t=0^+} <0$, and from below if $\left(d \delta j/dt
\right)_{t=0^+} >0$.

The relations between the frequency-dependent capacitance and transient
current in response to a voltage step are very general and applicable to any
type of electronic device, although the microscopic mechanism of the
transient response can be quite different.

To illustrate these considerations, let us consider a simple yet very general
type of transient response composed of negative and positive exponential
components:
\begin{equation}
\Delta j(t)=\Delta V \left\{a_1 \exp(-t/\tau_1)
- a_2 \exp(-t/\tau_2) \right\},
\label{anl-I}
\end{equation}

\noi where $a_1$, $a_2$, $\tau_1$ and $\tau_2$ are some parameters. The
capacitance corresponding to this transient is calculated using
eq.~(\ref{cap-EQ}):
\begin{equation}
C(\omega)=C_0+{{a_1\tau_1}\over{1+
(\omega\tau_1)^2}} - {{a_2\tau_2}\over{1+ (\omega\tau_2)^2}}.
\label{anl-C}
\end{equation}

Figure~4 shows the relaxation current and corresponding frequency-dependent
capacitance for different values of the parameters listed in the inset of
Fig.~4(a). It can be seen that, depending on values of the parameters, the
frequency dependence of capacitance can be either monotonic with $C>0$, or
non-monotonic (if the transient current is non-monotonic) with $C<0$ at low
frequency, in accordance with the general theoretical considerations stated
above.

It is interesting to note that relations similar to those considered in this
section are true for the frequency-dependent conductance and transient
current in response to an impulse-like voltage signal, which follows from the
similarity of eqs.(\ref{cap-EQ})--(\ref{cond-EQ}) and
(\ref{cap-EQA})--(\ref{cond-EQA}) (see also ref.~\cite{NegCond71}). Similar
relationships can also be obtained by analyzing the transient voltage
response to an excitation by a current impulse~\cite{Shockley54,MisawaNC57}.

\section{NC in quantum well infrared photodetectors}
\subsection{Device structure and physics}
The physics of QWIPs has been extensively discussed in the
literature~\cite{LevineR,ErshovJJAP96}. Fig.~5 shows the geometry and diagram
of the physical processes in QWIP. The QWIP structure comprises the QW region
including doped QWs separated by undoped barriers. The QW structure is clad
with the emitter and collector contacts. Under applied voltage, electrons are
injected from the emitter and drift towards the collector. Electrons can be
captured by the QWs and emitted from the QWs to the continuum due to
thermoexcitation (we consider dark current conditions here). The value of the
electric field at the injecting contact is controlled by the recharging of
the QWs near the emitter to equilibrate the injection current and drift
current in the bulk of QWIP. Our preliminary simulations of QWIPs showed that
the capacitance can be negative for some voltage and frequency ranges. This
result was surprising, since the NC phenomenon has never been reported in
QWIPs before, even though these devices have been actively studied over the
last decade. Therefore, we checked this result experimentally to exclude the
possibility of simulation artifacts.

\subsection{Experimental data}
The results presented here were obtained on a GaAs/Al$_{0.251}$Ga$_{0.749}$As
QWIP with 4 QWs of 62~$\AA$ width, separated by barriers of 241~$\AA$ width.
The barriers were undoped, and the QWs were center $\delta$-doped with
silicon to about 9$\times$10$^{11}$~cm$^{-2}$. The GaAs contacts doped to
1.5$\times$10$^{18}$~cm$^{-3}$ were separated from the QW structure by
rectangular barriers identical to inter-well barriers. The geometric
capacitance was $C_0=10.9$~pF for a mesa device of $120\times120$~$\mu$m$^2$
in size. All measurements were performed at a temperature of $T$=80~K.
Devices were mounted on a test package with equivalent open and short wires
for reference. HP4284A (20~Hz--1~MHz) and 4285A (75~kHz--30~MHz) Precision
LCR meters were used for the C-V and C-F measurements. We checked carefully
that parasitic elements did not influence the measurement data. Static
characteristics of this QWIP were reported in ref.~\cite{LiuNW92}.

Figure~6 shows C-V characteristics measured at different frequencies. For the
lowest frequency of 0.1~kHz the capacitance displays a maximum at zero bias.
With an increase of voltage the capacitance decreases rapidly, approaching
negative values. The capacitance does not decrease monotonically with voltage
but displays peaks and shoulders. The capacitance at zero bias decreases with
frequency, approaching the value of the geometric capacitance at frequencies
$f\ge 10^2$--$10^3$~kHz. C-V characteristics at intermediate frequencies
(1~kHz$\le f \le $100~kHz) are similar to low-frequency characteristics. The
decrease of capacitance with voltage becomes less steep, and the voltage at
which $C$=0 increases with frequency. At the highest measurement frequencies
$f\ge$1~MHz the capacitance is constant and equal to C$_0$ at low voltage,
and exhibits an overshoot at high voltages, with a subsequent steep decrease.
For this frequency range, the capacitance does not reach negative values at
negative bias, as it would require too high a voltage, causing device
heating.

The frequency dependence of the capacitance is shown in Fig.~7 using two
different vertical scales. The magnitudes of the NC values are plotted in the
log plot (Fig.~7(b)). At very low frequencies ($f\le 100$~Hz) the capacitance
data are very noisy and not plotted. The capacitance at low bias voltages
($|V|\le $0.1~V) is positive and decreases with frequency to the value of
$C_0$. For a higher voltage ($V=-0.2$~V), the capacitance starts with
negative values at low frequencies and increases monotonically above zero,
approaching the $C_0$ value at high frequencies. With the further increase of
voltage ($V=-0.49$~V) the capacitance dependence on frequency becomes
non-monotonic and develops a broad maximum. The magnitude and frequency
location of its maximum depend on the applied voltage. The absolute value of
the negative capacitance at low frequencies increases rapidly with voltage,
and can exceed the geometric capacitance by more than two orders of magnitude
(Fig.~7b).

Measurements on QWIPs having 8, 16 and 32 QWs gave a similar capacitance
behavior as a function on bias voltage and frequency, but the peaks on the
C-V curves were less pronounced.

\subsection{Simulation results}
We performed the simulation with the use of a time-dependent QWIP simulator
based on the numerical model described in ref.~\cite{ErshovAPL95}. The
capacitance was calculated using the Fourier transform of the transient
current (see eq.~(\ref{cap-EQ})) obtained from the solution of the
time-dependent problem on application of a small voltage step ($\Delta
V$=5$\times10^{-3}$~V) to the QWIP. Calculated C-V and C-F characteristics
are shown in Fig.~8. We would like to point out the very good qualitative
agreement of the simulation results and experimental data by comparing Fig.~8
with Figs.~6 and 7. However, there are some quantitative discrepancies,
including the magnitude of the low-frequency capacitance at low voltage, the
width and magnitude of the capacitance peaks in the C-V and C-F
characteristics, the magnitude of the negative capacitance, and the
frequencies at which C=0. In our computational experiments we found that
these features are {\em very} sensitive to the parameters of the simulation
model (such as the QW capture velocity, field-dependent mobility, escape time
from the QWs, etc.) and on the operating temperature. However, the main
features of the capacitance behavior are independent of the variation of
model parameters and temperature: at low voltages capacitance at low
frequency is positive and decreases to the geometric capacitance at high
frequency; capacitance at high voltages is negative at low frequency and
increases with frequency to reach the geometric capacitance; the magnitude of
the negative capacitance strongly increases with voltage and can exceed the
geometric capacitance by a few orders of magnitude; and the voltage at which
C=0 increases with increasing frequency. Since the model parameters are not
available, and taking into account a significant asymmetry of experimental
C-V characteristics, we focused not on the fitting of simulation results to
experimental data, but on the explanation of the unusual features of the QWIP
capacitance and the underlying physics.

As we discussed above, the clue to the capacitance behavior in the frequency
domain is in the time-domain transient current. Figure~9 shows the calculated
transient current for several applied voltages. The relaxation component of
the transient current is positive and monotonic at low voltages, which
results in positive capacitance at any frequency in the low voltage range
(Fig.~8(a)). At high voltages, the transient current is dominated by a
negative component, whose amplitude increases with voltage (Fig.~8(b)). Since
the low-frequency capacitance is directly related to the transient current,
this results in negative low-frequency capacitance, which increases strongly
in magnitude with bias. The high-frequency capacitance tends to the geometric
capacitance. Physically, this is due to the fact that the QW recharging
processes determining the capacitance are ``frozen'' at frequencies higher
than the inverse characteristic time of QW recharging.

We now give a physical picture of the transient current behavior. The
transient current after the application of a voltage step is due to several
effects resulting from an instantaneous increase of the electric field, such
as enhanced electron emission from the QWs, increased drift electron
velocity, and non-equilibrium electron injection from the emitter. The
combined influence of these effects on transient current can be quite
complicated in general, and their relative importance depends strongly on
applied voltage. At high applied voltages ($|V|\ge 0.2$~V for the present
4-well sample) the effect of the non-equilibrium transient injection is the
dominant one in determining the behavior of the transient current, including
negative current $\delta j(t)<0$. To illustrate this effect, we plot in
Fig.~10 the dependence of the electric field in the emitter barrier $E_e$ on
the average electric field $E_a(V)=V/L$ along with its derivative
$dE_e/dE_a(V)$ under steady-state conditions. It is seen that $dE_e/dE_a>1$
at any $E_a$ for the QWIP under consideration. This effect results from the
difference of the current-electric field characteristics in the injecting
contact and the bulk QW region.~\cite{ErshovJJAP96} Note that just after the
application of a voltage step $\Delta V$ to the QWIP, the instantaneous
change of the electric field is constant over the QWIP structure and equal
to $\Delta E=\Delta V/L$. This means that at the beginning of the transient,
the electric field in the emitter barrier $E_e(V)+\Delta E$ is {\em lower}
than the steady-state electric field $E_e(V+\Delta V)$ corresponding to the
new voltage $V+\Delta V$ (see Fig.~11). Therefore, the injection
(conduction) current is {\em lower} than the steady-state current, resulting
in a negative transient current $\delta j$. The transient current at high
voltage is dominated by the conduction component. The magnitude of the
negative transient current at time $t=0^+$ is equal approximately to
$d\,j_{inj}/dE_e\times [E_e(V+\Delta V)-(E_e(V)+\Delta E)]$, and increases
strongly with voltage due to an exponential dependence of injection current
$j_{inj} (E_e)$ ~\cite{ErshovJJAP96}. During the transient, the QWs are
recharged and the current tends to its steady-state value. Thus, the
non-monotonic and negative-valued behavior of $\delta j$, responsible for
the negative capacitance, are due to the non-equilibrium transient injection
from the emitter upon application of a voltage step.

At low voltage, the conductance of the QWIP is low, and the transient
current is dominated by the displacement component related to the
release and escape of electrons from the QWs.

It should be noted that transient current characteristics are determined by
the $E_e(E_a)$ behavior, and therefore, by the injecting property of the
emitter. Depending on the structural parameters of the injecting barrier, the
derivative $dE_e/dE_a(V)$ can be either greater or less than unity, thus
making the transient current negative or positive. Our simulations show that
QWIPs with a triangular emitter barrier~\cite{ErshovNGS95} have
$dE_e/dE_a<1$. The transient current in these QWIPs is positive, and their
capacitance is always positive at all frequencies and voltages. In this
respect, QWIPs with triangular emitter and multiple QWs are similar to the
single QW phototransistor considered
earlier~\cite{Ryzhii_S_JAP95,ErshovIEEE96}.

Thus, the negative-valued behavior of the transient current in the QWIP in
response to a voltage step and the NC is due to the non-equilibrium
transient injection from the emitter.

We have to point out that some features of the experimental data on QWIP
capacitance are still unexplained. Simulation predicts the saturation of NC
at low frequencies (Fig.~8(b)), while experimental data shows a strong
increase of the absolute value of NC proportional to $1/\omega$ at low
frequencies (Fig.~7(b)). A similar behavior of NC has also been reported in
other semiconductor devices~\cite{JonscherNC88,JonscherURL}, but the physical
mechanism of this low-frequency behavior of QWIP capacitance is not clear.
This phenomenon confirms a close relation between the NC effect and the
effect of the low-frequency dispersion observed in a great variety of
semiconductor devices and other physical systems~\cite{JonscherURL}.

\section{Discussion}
A negative capacitance $C$ has the same phase relationship between the
small-signal voltage and current as a positive inductance $L=-1/\omega^2C$.
However, the interpretation of negative capacitance in terms of conventional
inductance is not physically meaningful for the following reasons. First, in
the case of ``normal'' inductance the behavior is associated with the
magnetic field, which is not relevant for our consideration. Second, the
impedance of ``normal'' inductance $|Z|=\omega L$ increases with frequency
and, therefore, should dominate at high frequencies. However, the
semiconductor devices under consideration display normal capacitance behavior
($C=C_0>$0) at high frequency. In this regard, the interpretation of the NC
effect in terms of parasitic series inductances~\cite{Butcher96} or poor
measurement equipment calibration~\cite{Huang97} is incorrect. If, to the
contrary, high-frequency capacitance is negative, and the imaginary part of
the impedance is proportional to the frequency $|Z|\sim \omega$, this is a
clear indication of the dominant role of parasitic inductance of the external
circuit (see, for example, ref.~\cite{MPekoNC97}). Thus, measurement of the
device admittance (or impedance) in a wide frequency range, and its behavior
at high frequencies tests whether the NC effect is due to internal properties
of the device or due to the parasitic inductance.

A term ``negative capacitance'' is often used to emphasize the inductive
behavior of a device, which is expected to display (or usually displays)
capacitive response. Complementary, a term ``negative inductance'' is used to
emphasize capacitive behavior of some devices (see, for example,
ref.~\cite{NegInd1}).

A note should be made concerning the physical interpretation of capacitance
and the methods of its calculation in semiconductor devices. Conventionally,
capacitance is associated with the accumulation of charges and electric field
energy with the change of the voltage on contacts. This concept comes from
electrostatics, when the conduction current is zero and the total electric
current is due to the displacement component, related to the redistribution
of charges inside the structure. However, the capacitance is determined by
the reactive part of the {\em total} current, which comprises both conduction
and displacement components. To distinguish the notion of ``true''
capacitance and ``electrostatic'' capacitance, a new term ``emittance'' has
been introduced recently~\cite{ButtPRL96}. These two concepts are equivalent
only if the conduction component of the reactive current is much lower than
the displacement component. Indeed, let us suppose that the conduction
component of the transient current $\Delta I(t)$ is zero at some
cross-section (in the 1D geometry), i.~e. the transient current contains only
the displacement component $\Delta I(t)= A\eps\eps_0
\partial E/\partial t$. Substituting this expression into
formula~(\ref{cap-eq}), we obtain (for the case $\omega=0$)
$C(0)=A\eps\eps_0\Delta E/\Delta V=\Delta Q/\Delta V$, where $\Delta Q$ is
the charge increment at the each side of that cross-section. Therefore, the
incremental charge approach to the capacitance calculation (for the case
$\omega=0$) is correct, if there is a cross-section inside the device where
the conduction current is zero (in devices with multi-dimensional geometry
(2D or 3D) and many contacts, low-frequency capacitance of a contact can be
rigorously calculated using this approach if the contact is non-conducting in
DC regime).

If the contribution of the conduction current to the reactive current is
larger than that of the displacement current (this is usually the case for
devices displaying NC), the capacitance is determined by the current which
passes through the structure without charging effects, and capacitance has no
relation to charge or energy accumulation. In this case, approaches of
capacitance calculation based on static device characteristics (such as the
incremental charge partitioning approach~\cite {LauxSS85}) are incorrect, and
the rigorous methods such as Fourier decomposition of transient excitation or
SSSA~\cite {LauxSS85} should be used.

It is worthwhile to mention a confusion caused by two different conventions
for defining the complex phase factor for small-signal harmonic quantities
(current, voltage, etc.). In this paper we followed the electrical (+)
convention $\delta I, \delta V \sim \exp(i\omega t)$. In this convention,
{\it capacitive} response (positive capacitance) corresponds to {\it
positive} reactive part of admittance (susceptance). The physics ($-$)
convention uses the phase factor $\exp(-i\omega t)$, which corresponds to
{\it capacitive} response if susceptance is {\it negative} (capacitance is
defined as $C=-{1\over\omega} \rm{Im} \left[ Y(\omega) \right]$ in physics
convention). Note that admittances (eq.~(\ref{imp-def})) corresponding to (+)
and ($-$) conventions are complex conjugate quantities: $Y_+(\omega) =
Y_-(\omega)$. However, important physical quantities (which can be measured
or calculated) and relationships (such as phase relationship between current
and voltage) are independent of the choice of the sign convention.

\section{Conclusions}
The effect of NC in semiconductor devices is discussed. The relations between
the transient relaxation current in the time-domain in response to a voltage
step or impulse and capacitance in the frequency-domain are outlined. NC
appears if the time-derivative of the transient current in response to a
voltage step is positive-valued or non-monotonic with time. The incremental
charge method of capacitance calculation is {\it absolutely inapplicable} in
the case of large conduction current in the device, which is often the case
when the capacitance is negative. The correct interpretation of NC should be
based on rigorous approaches such as SSSA or Fourier analysis of the
transient current. These points are illustrated by the results of
experimental and theoretical studies of small-signal characteristics of
QWIPs, which exhibit a huge NC. NC and peaks on the C-V and C-F curves are
explained in terms of the transient current, which has a positive-valued
time-derivative of transient current in the time-domain. This behavior is due
to the non-equilibrium transient electron injection from the emitter, which
is determined by the inertia of the QWs' recharging processes and injection
properties of the emitter barrier.

\section*{Acknowledgments}
This work has been partially supported by Electronic Communication Frontier
Research and Development Grant of Ministry of Post and Telecommunications,
Japan, by the Research Fund of the University of Aizu, and by Defence
Research Establishment Valcartier, Department of National Defence, Canada. We
thank P.~Chow-Chong and P.~Marshall of NRC for sample fabrication. One of the
authors (M.~E.) thanks M.~B\"uttiker, M.~J.~Morant, and M.~Stockman for
useful discussions.

\newpage
%\bibliographystyle{IEEE}
%\bibliography{abbrvm,qwipn}

\begin{thebibliography}{10}

\bibitem{KanaiNC55}
Y.~Kanai,
\newblock ``On the inductive part of the a.~c. characteristics of the
  semiconductor diodes,''
\newblock {\em J. Phys. Soc. Japan}, vol. 10, pp. 718--720, 1955.

\bibitem{MisawaNC57}
T.~Misawa,
\newblock ``Impedance of bulk semiconductor in junction diode,''
\newblock {\em J. Phys. Soc. Japan}, vol. 12, pp. 882--890, 1957.

\bibitem{MisawaNR66a}
T.~Misawa,
\newblock ``Negative resistance in p-n junctions under avalanch breakdown
  conditions, {Part I},''
\newblock {\em IEEE Trans. Electron Devices}, vol. 13, no. 1, pp. 137--141,
  1966.

\bibitem{MisawaNR66b}
T.~Misawa,
\newblock ``Negative resistance in p-n junctions under avalanch breakdown
  conditions, {Part II},''
\newblock {\em IEEE Trans. Electron Devices}, vol. 13, no. 1, pp. 141--153,
  1966.

\bibitem{KazarinovNC67}
R.~F. Kazarinov, V.~I. Stafeev, and R.~A. Suris,
\newblock ``Impedance and transient processes in germanium diodes with deep
  levels,''
\newblock {\em Sov. Phys. Semicond.}, vol. 1, no. 9, pp. 1084--1086, 1967.

\bibitem{MarshNC67}
O.~J. Marsh and C.~R. Viswanathan,
\newblock ``Space-charge-limited current of holes in silicon and techniques for
  distinguishing double and single injection,''
\newblock {\em J. Appl. Phys.}, vol. 38, no. 8, pp. 3135--3144, 1967.

\bibitem{VogelNC69}
R.~Vogel and P.~J. Walsh,
\newblock ``Negative capacitance in amorphous semiconductor chalcogenide thin
  films,''
\newblock {\em Appl. Phys. Lett.}, vol. 14, no. 7, pp. 216--218, 1969.

\bibitem{RockstadNC71}
H.~K. Rockstad,
\newblock ``Ionization model for negative capacitance in low-mobility
  semiconductors such as amorphous chalcogenides,''
\newblock {\em J. Appl. Phys.}, vol. 42, no. 3, pp. 1159--1166, 1971.

\bibitem{AllisonNC71}
J.~Allison and V.~R. Dave,
\newblock ``Frequency dependence of negative-capacitance effects observed in
  amorphous semiconductor thin-film devices,''
\newblock {\em Electron. Lett.}, vol. 7, no. 24, pp. 706--707, 1971.

\bibitem{AltunyanNC71}
S.~A. Altunyan, V.~S. Minaev, V.~I. Stafeev, L.~S. Gasanov, A.~S. Deshevoi, and
  B.~K. Skachkov,
\newblock ``An investigation of the electrical characteristics of memory and
  nonmemory switching devices made of chalcogenide glasses,''
\newblock {\em Sov. Phys. Semicond.}, vol. 5, no. 3, pp. 427--430, 1971.

\bibitem{BrodovoiNC73}
V.~A. Brodovoi and N.~Z. Derikot,
\newblock ``Investigation of the impedance of {GaAs:Cu} in strong electric
  fields,''
\newblock {\em Sov. Phys. Semicond.}, vol. 7, no. 4, pp. 459--461, 1973.

\bibitem{KolomietsNC74}
B.~T. Kolomiets, A.~Ya. Karachentsev, and V.~V. Spevak,
\newblock ``Investigation of surface barriers in silicon carbide single
  crystals,''
\newblock {\em Sov. Phys. Semicond.}, vol. 7, no. 7, pp. 872--875, 1974.

\bibitem{EgiazaryanNC75}
G.~A. Egiazaryan and V.~I. Stafeev,
\newblock ``Some properties of {S}-type dioes made of semiinsulating gallium
  arsenide,''
\newblock {\em Sov. Phys. Semicond.}, vol. 9, no. 3, pp. 334--336, 1975.

\bibitem{DeshevoiNC77}
A.~S. Deshevoi and L.~S. Gasanov,
\newblock ``Solid-state inductance of amorphous and compensated
  semiconductors,''
\newblock {\em Sov. Phys. Semicond.}, vol. 11, no. 10, pp. 1168--1170, 1977.

\bibitem{VeingerNC78}
A.~I. Veinger,
\newblock ``Contact with a negative differential capacitance in the form of
  hot-carrier p-n junction,''
\newblock {\em Sov. Phys. Semicond.}, vol. 12, no. 10, pp. 1180--1182, 1978.

\bibitem{NoguchiNC80}
T.~Noguchi, M.~Kitagawa, and I.~Taniguchi,
\newblock ``Negative capacitance of silicon diode with deep level traps,''
\newblock {\em Jpn. J. Appl. Phys.}, vol. 19, no. 7, pp. 1423--1424, 1980.

\bibitem{NadkarniNC83}
G.~S. Nadkarni, N.~Sankarraman, and S.~Radhakrishnan,
\newblock ``Switching and negative capacitance in
  {Al-Ge$_{15}$Te$_{81}$Sb$_2$S$_2$-Al} devices,''
\newblock {\em J. Phys. D}, vol. 16, pp. 897--908, 1983.

\bibitem{AlimpievNC84}
V.~N. Alimpiev and I.~R. Gural'nik,
\newblock ``Negative capacitance of a photosensitive semiconductor,''
\newblock {\em Sov. Phys. Semicond.}, vol. 18, no. 4, pp. 420--422, 1984.

\bibitem{BlatterNC86}
G.~Blatter and F.~Greuter,
\newblock ``Electrical breakdown at semiconductor grain boundaries,''
\newblock {\em Phys. Rev. B}, vol. 34, no. 12, pp. 8555--8572, 1986.

\bibitem{JonscherNC86}
A.~K. Jonscher, C.~Pickup, and S.~Zaidi,
\newblock ``Dielectric spectroscopy of semi-insulating gallium arsenide,''
\newblock {\em Semicond. Sci. Technol.}, vol. 1, pp. 71--92, 1986.

\bibitem{FuNC87}
S.-T. Fu and M.~B. Das,
\newblock ``Backgate-induced characteristics of ion-implanted {GaAs}
  {MESFET's},''
\newblock {\em IEEE Trans. Electron Devices}, vol. 34, no. 6, pp. 1245--1252,
  1987.

\bibitem{JonscherNC88}
A.~K. Jonscher and M.~N. Robinson,
\newblock ``Dielectric spectroscopy of silicon barrier devices,''
\newblock {\em Solid-State Electron.}, vol. 31, no. 8, pp. 1277--1288, 1988.

\bibitem{WernerNC88}
J.~Werner, A.~F.~J. Levi, R.~T. Tung, M.~Anzlowar, and M.~Pinto,
\newblock ``Origin of the excess capacitance at intimate {Schottky} contacts,''
\newblock {\em Phys. Rev. Lett.}, vol. 60, no. 1, pp. 53--56, 1988.

\bibitem{WuNC89}
X.~Wu and E.~S. Yang,
\newblock ``Interface capacitance in metal-semiconductor junctions,''
\newblock {\em J. Appl. Phys.}, vol. 65, no. 9, pp. 3560--3567, 1989.

\bibitem{HeNC89}
L.~He and T.~Dingyuan,
\newblock ``Capacitance-voltage characteristics of p-n junction of the narrow
  band-gap semiconductors {Hg$_{1-x}$Cd$_{x}$Te},''
\newblock {\em Chinese Physics}, vol. 9, no. 1, pp. 223--230, 1989.

\bibitem{WuNC90}
X.~Wu, H.~L. Ebans, and E.~S. Yang,
\newblock ``Negative capacitance at metal-semiconductor interfaces,''
\newblock {\em J. Appl. Phys.}, vol. 68, no. 6, pp. 2845--2848, 1990.

\bibitem{ChampnessNC90}
C.~H. Champness and W.~R. Clark,
\newblock ``Anomalous inductive effect in selenium {Schottky} diodes,''
\newblock {\em Appl. Phys. Lett.}, vol. 56, no. 12, pp. 1104--1106, 1990.

\bibitem{MerlinNC90}
R.~Merlin and D.~A. Kessler,
\newblock ``Photoexcited quantum wells: Nonlinear screening, bistability, and
  negative differential capacitance,''
\newblock {\em Phys. Rev. B}, vol. 41, no. 14, pp. 9953--9957, 1990.

\bibitem{MuretNC91}
P.~Muret, D.~Elguennouni, M.~Missous, and E.~H. Rhoderick,
\newblock ``Admittance of {Al/GaAs} {Schottky} contacts under forward bias as a
  function of interface preparation conditions,''
\newblock {\em Appl. Phys. Lett.}, vol. 58, no. 2, pp. 155--157, 1991.

\bibitem{MorantNC92}
M.~J. Morant and B.~Y. Majlis,
\newblock ``A silicon negative resistance, negative capacitance device,''
\newblock in {\em Proc. of the IEEE Int. Conf. on Semiconductor Electronics},
  1992.

\bibitem{BealeNC92a}
M.~Beale and P.~Mackay,
\newblock ``The origins and characteristics of negative capacitance in
  metal-insulator-metal devices,''
\newblock {\em Philosophical Magazine B}, vol. 65, no. 1, pp. 47--64, 1992.

\bibitem{BealeNC92b}
M.~Beale,
\newblock ``Anomalous reactance behaviour during the impedance analysis of
  time-varying dielectric systems,''
\newblock {\em Philosophical Magazine B}, vol. 65, no. 1, pp. 65--77, 1992.

\bibitem{BoltaevNC95}
A.~P. Boltaev, T.~M. Burbaev, G.~A. Kalyuzhnaya, V.~A. Kurbatov, and N.~A.
  Penin,
\newblock ``Negative capacitance in {Ni-TiO$_2$-p-Si} heterostructures,''
\newblock {\em Russian Microelectronics}, vol. 24, no. 4, pp. 255--258, 1995.

\bibitem{ButtPRL96}
T.~Christen and M.~B\"uttiker,
\newblock ``Low frequency admittance of a quantum point contact,''
\newblock {\em Phys. Rev. Lett.}, vol. 77, no. 1, pp. 143--146, 1996.

\bibitem{KuttyNC96}
S.~Ezhilvalavan and T.~R.~N. Kutty,
\newblock ``High-frequency capacitance resonance of {ZnO}-based varistor
  ceramics,''
\newblock {\em Appl. Phys. Lett.}, vol. 69, no. 23, pp. 3540--3542, 1996.

\bibitem{ErshovCapAPL}
M.~Ershov, H.~C. Liu, L.~Li, M.~Buchanan, Z.~R. Wasilewski, and V.~Ryzhii,
\newblock ``Unusual capacitance behavior of quantum well infrared
  photodetectors,''
\newblock {\em Appl. Phys. Lett.}, vol. 70, no. 14, pp. 1828--1830, 1997.

\bibitem{MisiakosNC97}
K.~Misiakos, D.~Tsamakis, and E.~Tsoi,
\newblock ``Measurement and modeling of the anomalous dynamic response of high
  resistivity diodes at cryogenic temperatures,''
\newblock {\em Solid-State Electron.}, vol. 41, no. 8, pp. 1099--1103, 1997.

\bibitem{OmuraNC97}
I.~Omura, H.~Ohashi, and W.~Fichtner,
\newblock ``{IGBT} negative gate capacitance and related instability effects,''
\newblock {\em IEEE Electron Device Lett.}, vol. 18, no. 12, pp. 622--624,
  1997.

\bibitem{Jonscher86}
A.~K. Jonscher,
\newblock ``The physical origin of negative capacitance,''
\newblock {\em J.~Chem. Soc., Faraday Trans. II}, vol. 82, pp. 75--81, 1986.

\bibitem{JonscherURL}
A.~K. Jonscher,
\newblock {\em Universal Relaxation Law},
\newblock Chelsea Dielectrics Press, London, 1996.

\bibitem{Butcher96}
K.~S.~A. Butcher, T.~L. Tansley, and D.~Alexiev,
\newblock ``An instrumental solution to the phenomenon of negative capacitances
  in semiconductors,''
\newblock {\em Solid-State Electron.}, vol. 39, no. 3, pp. 333--336, 1996.

\bibitem{Huang97}
X.~L. Huang, Y.~G. Shin, K.~Y. Lim, E.-K. Suh, H.~J. Lee, and S.~C. Shen,
\newblock ``Thermally induced capacitance and electric field domains in
  {GaAs/Al$_{0.3}$Ga$_{0.7}$As} quantum well infrared photodetectors,''
\newblock {\em Solid-State Electron.}, vol. 41, no. 6, pp. 845--850, 1997.

\bibitem{Negcap96}
N.~A. Penin,
\newblock ``Negative capacitance in semiconductor structures,''
\newblock {\em Semiconductors}, vol. 30, no. 4, pp. 340--343, 1996.

\bibitem{LauxSS85}
S.~E. Laux,
\newblock ``Techniques for small-signal analysis of semiconductor devices,''
\newblock {\em IEEE Trans. Comput.-Aided Design Integrated Circuits}, vol. 4,
  no. 4, pp. 472--481, 1985.

\bibitem{NegCond71}
J.~C. McGroddy and P.~Gu\'eret,
\newblock ``Dynamic bulk negative differential conductivity in
  semiconductors,''
\newblock {\em Solid-State Electron.}, vol. 14, pp. 1219--1224, 1971.

\bibitem{Shockley54}
W.~Shockley,
\newblock ``Negative resistance arising from transit time in semiconductor
  devices,''
\newblock {\em The Bell System Technical Journal}, vol. 33, no. 4, pp.
  799--826, 1954.

\bibitem{LevineR}
B.~F. Levine,
\newblock ``Quantum-well infrared photodetectors,''
\newblock {\em J. Appl. Phys.}, vol. 74, no. 8, pp. R1--R81, 1993.

\bibitem{ErshovJJAP96}
M.~Ershov, C.~Hamaguchi, and V.~Ryzhii,
\newblock ``Device physics and modeling of multiple quantum well infrared
  photodetectors,''
\newblock {\em Jpn. J. Appl. Phys.}, vol. 35, Part 1, no. 2B, pp. 1395--1400,
  1996.

\bibitem{LiuNW92}
A.~G. Steele, H.~C. Liu, M.~Buchanan, and Z.~R. Wasilewski,
\newblock ``Influence of the number of wells on the performance of multiple
  quantum well intersubband infrared detectors,''
\newblock {\em J. Appl. Phys.}, vol. 72, no. 3, pp. 1062--1064, 1992.

\bibitem{ErshovAPL95}
M.~Ershov, V.~Ryzhii, and C.~Hamaguchi,
\newblock ``Contact and distributed effects in quantum well infrared
  photodetectors,''
\newblock {\em Appl. Phys. Lett.}, vol. 67, no. 21, pp. 3147--3149, 1995.

\bibitem{ErshovNGS95}
M.~Ershov and V.~Ryzhii,
\newblock ``Modeling of multiple {InGaAs/GaAs} quantum well infrared
  photodetectors,''
\newblock in {\em Proceedings of the 7th International Conference on Narrow Gap
  Semiconductors}, J.~L. Reno, Ed. Santa-Fe, NM, January 8--12, 1995, pp.
  353--358, IOP, Bristol, 1995.

\bibitem{Ryzhii_S_JAP95}
V.~Ryzhii and M.~Ershov,
\newblock ``Electron density modulation effect in a quantum-well infrared
  phototransistor,''
\newblock {\em J. Appl. Phys.}, vol. 78, no. 2, pp. 1214--1218, 1995.

\bibitem{ErshovIEEE96}
M.~Ershov, V.~Ryzhii, and K.~Saito,
\newblock ``Small-signal performance of a quantum well diode,''
\newblock {\em IEEE Trans. Electron Devices}, vol. 43, no. 3, pp. 467--472,
  1996.

\bibitem{MPekoNC97}
J.-C. M'Peko,
\newblock ``Effect of negative capacitances on high-temperature dielectric
  measurements at relatively low frequency,''
\newblock {\em Appl. Phys. Lett.}, vol. 71, no. 25, pp. 3730--3732, 1997.

\bibitem{NegInd1}
R.~Rifkin and Jr. B.~S.~Deaver,
\newblock ``Current-phase relation and phase-dependent conductance of
  superconducting point contacts from {RF } impedance measurements,''
\newblock {\em Phys. Rev. B}, vol. 13, no. 9, pp. 3894--3901, 1976.

\end{thebibliography}

%\newpage

\section*{Figure Captions}
\begin{enumerate}

\item[Fig. 1.]
Schematic diagram of (a) transient voltage and (b) transient current in a
device excited by a voltage step. The transient current is decomposed into
the components corresponding to (c) DC conductivity, (d) displacement
current, and (e) relaxation current.

\item[Fig. 2.]
Schematic diagram of (a) transient voltage and (b) transient current in a
device excited by a voltage impulse. The transient current is decomposed
into the components corresponding to (c) DC conductivity, (d) displacement
current, and (e) relaxation current.

\item[Fig. 3.]
Illustration of the relation between the transient relaxation current and
capacitance (eq.~(\ref{cap-EQA})) for the case of monotonically decreasing
function $[-d\delta j(t)/dt]$ ($C(\omega)-C_0>0$).

\item[Fig. 4.]
(a) Transient relaxation current and (b) capacitance for the analytical
model (equations (\ref{anl-I})--(\ref{anl-C})). The values of the parameters
are listed in the inset.

\item[Fig. 5.]
(a) Schematic diagram of structure and (b) conduction band profile for a
QWIP.

\item[Fig. 6.]
Capacitance-voltage characteristics measured at different frequencies. The
arrow indicates the value of the geometric capacitance $C_0$.

\item[Fig. 7.]
Frequency dependence of capacitance on (a) a linear scale and (b) a
logarithmic scale for different applied voltages. In (b), absolute value of
capacitance is plotted, and the parts of the curves corresponding to
negative capacitance values are indicated by arrows.

\item[Fig. 8.]
Calculated capacitance-voltage (a) and capacitance-frequency (b)
characteristics. In (b), absolute value of capacitance is plotted, and parts of
the curves corresponding to negative capacitance values are indicated by arrows.

\item[Fig. 9.]
Calculated transient current in response to a voltage step of $\Delta
V=5\times 10^{-3}$~V at (a) low and (b) high voltages.

\item[Fig. 10.]
Emitter electric field $E_e$ and its derivative versus
average electric field $E_a=V/L$.

\item[Fig. 11.]
Time dependence of (a) the electric field in the emitter barrier and (b) transient
current in QWIP in response to a voltage step.

\end{enumerate}

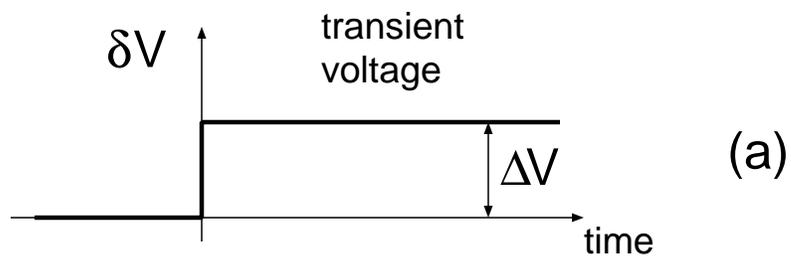
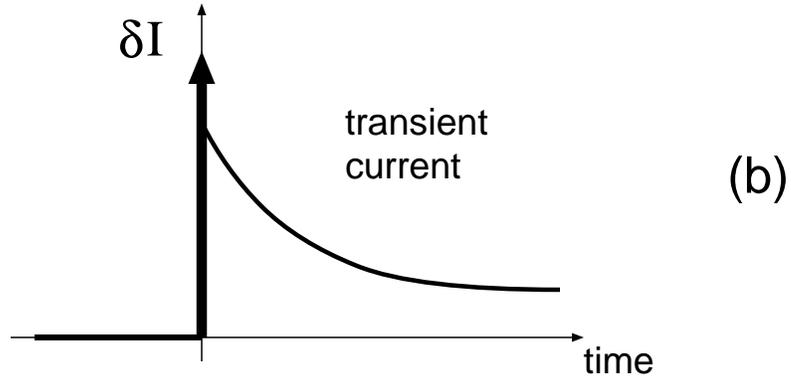
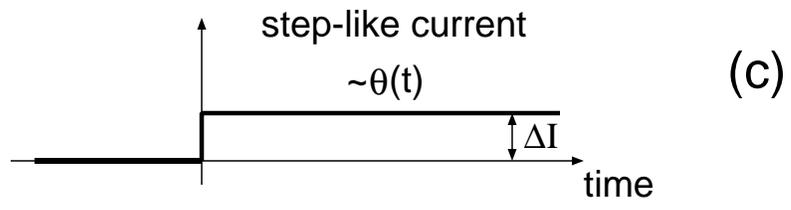
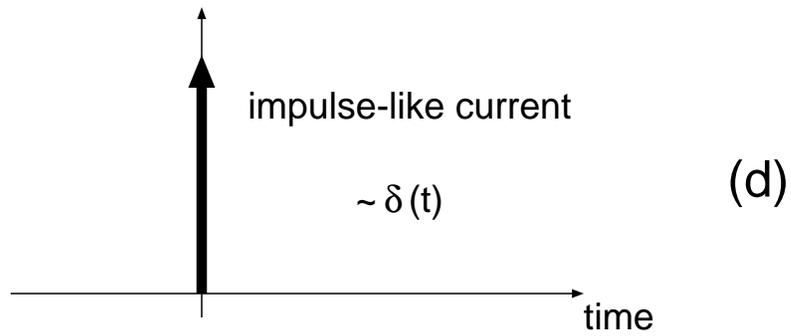
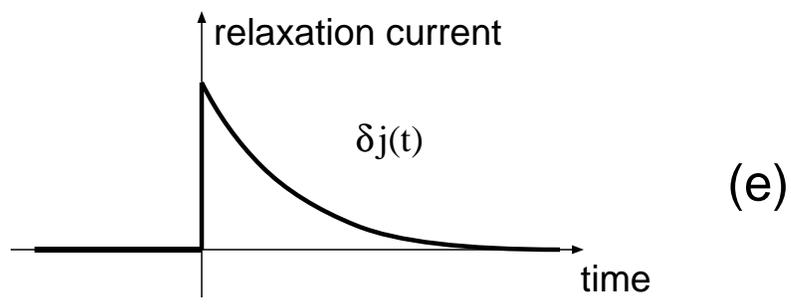
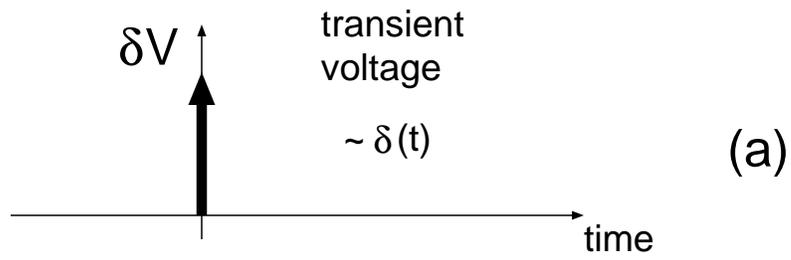
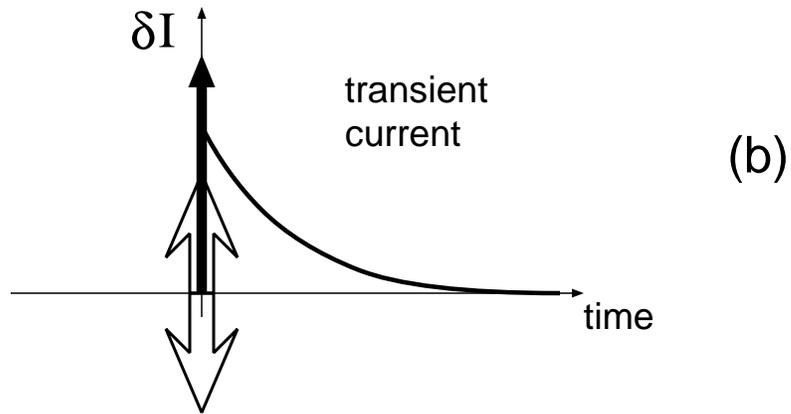
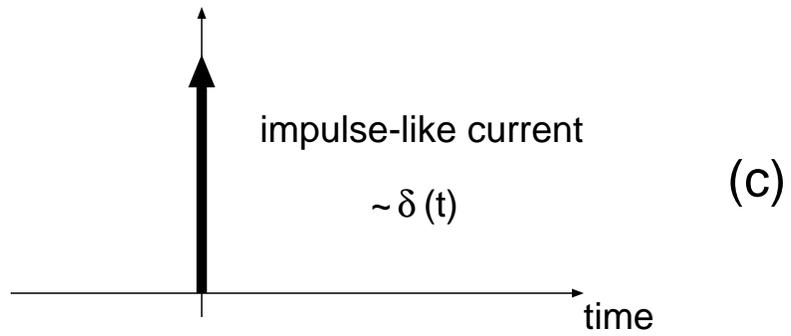
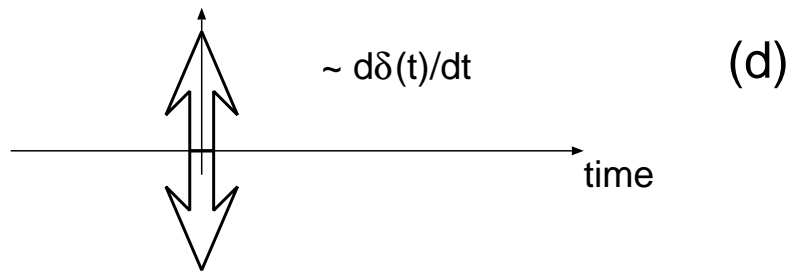
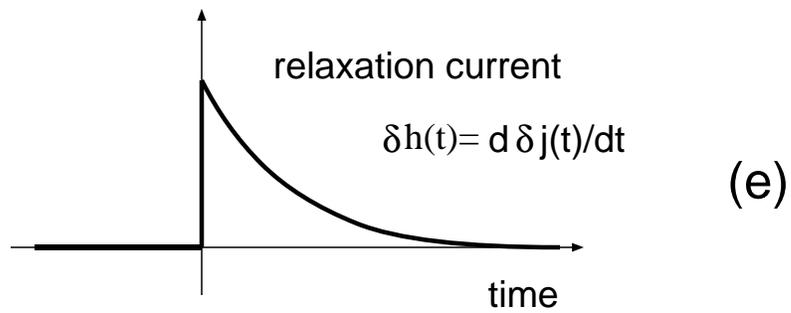
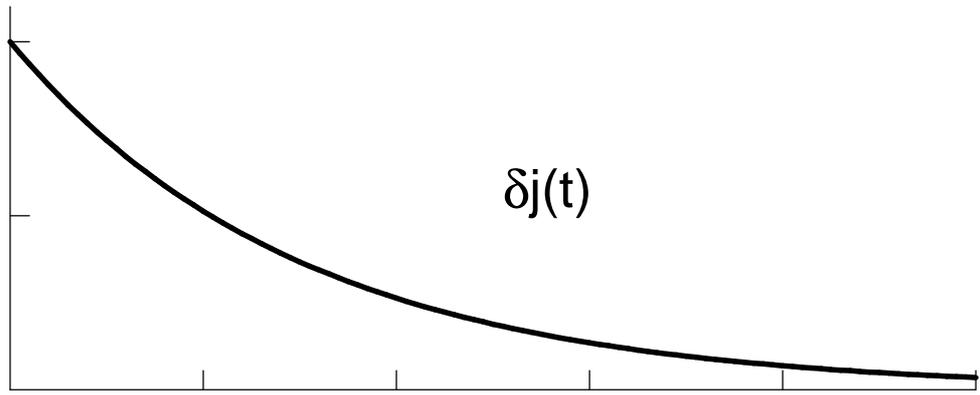
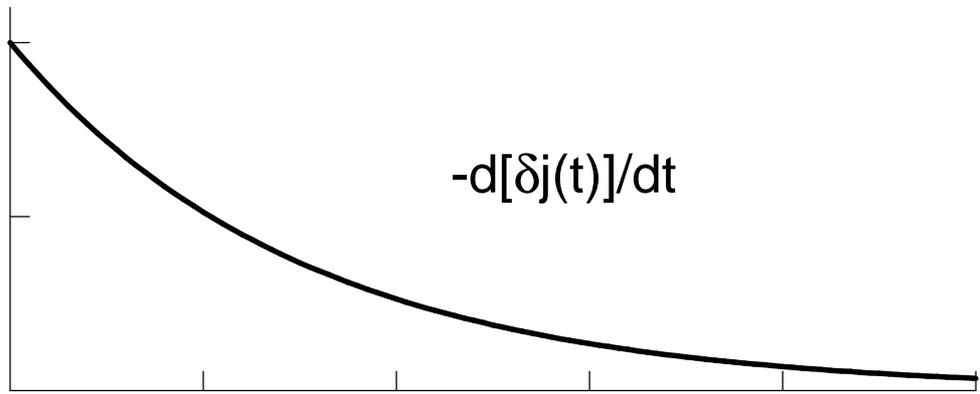
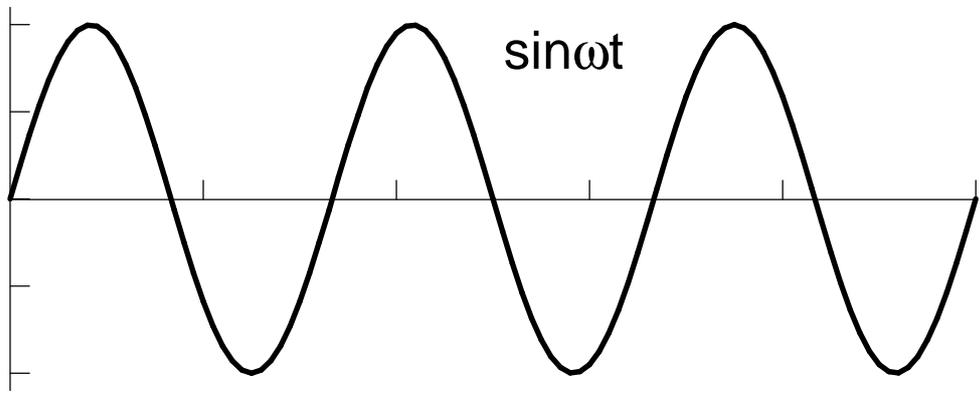
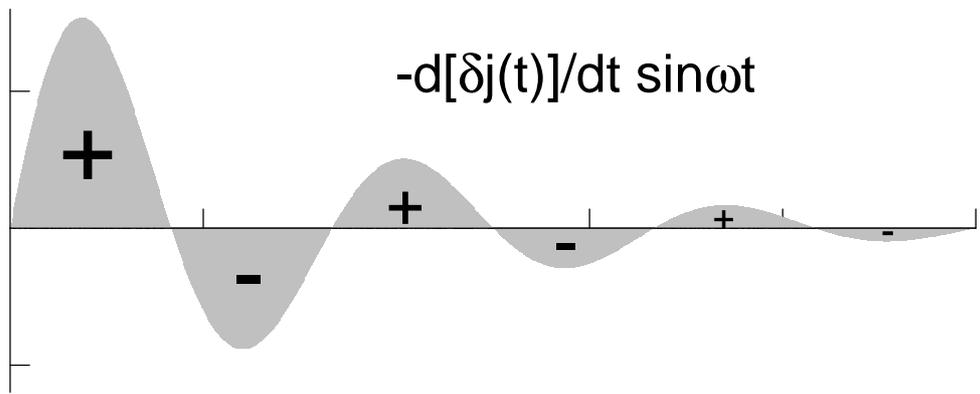
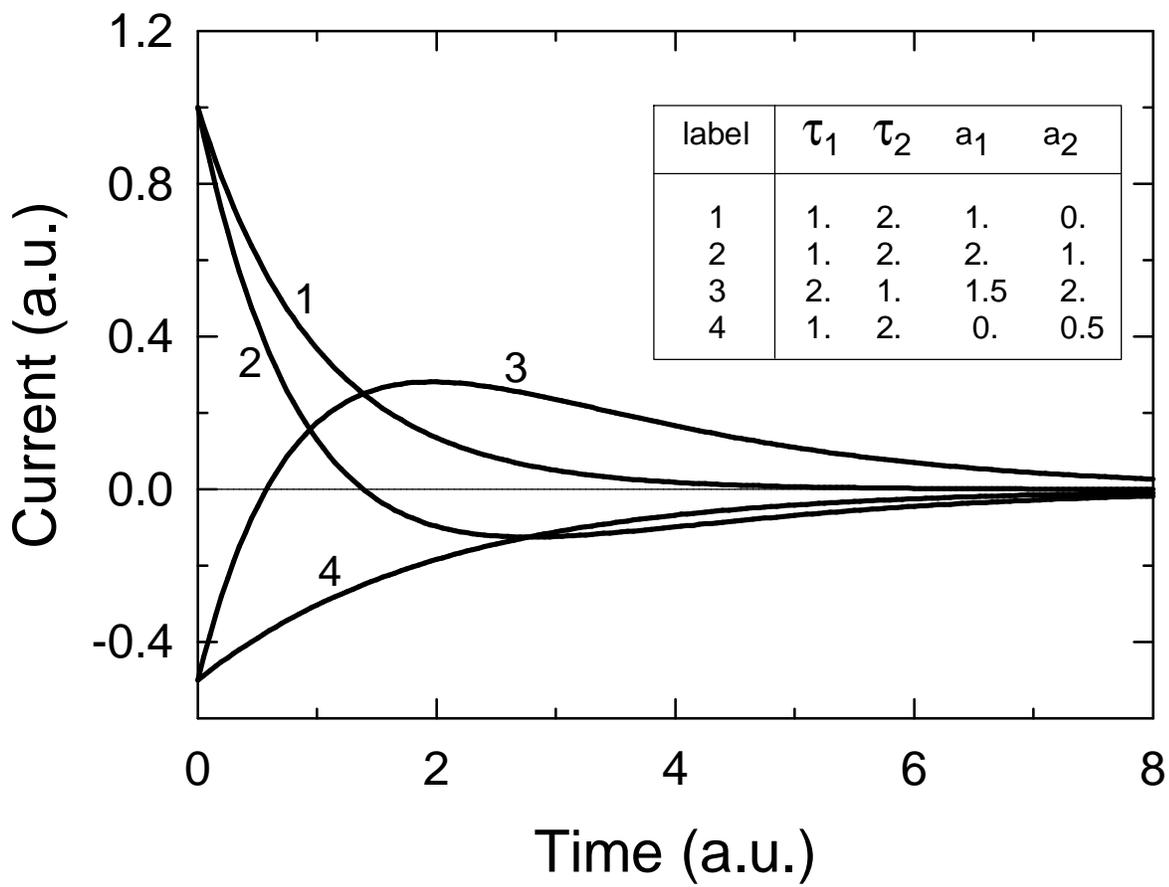
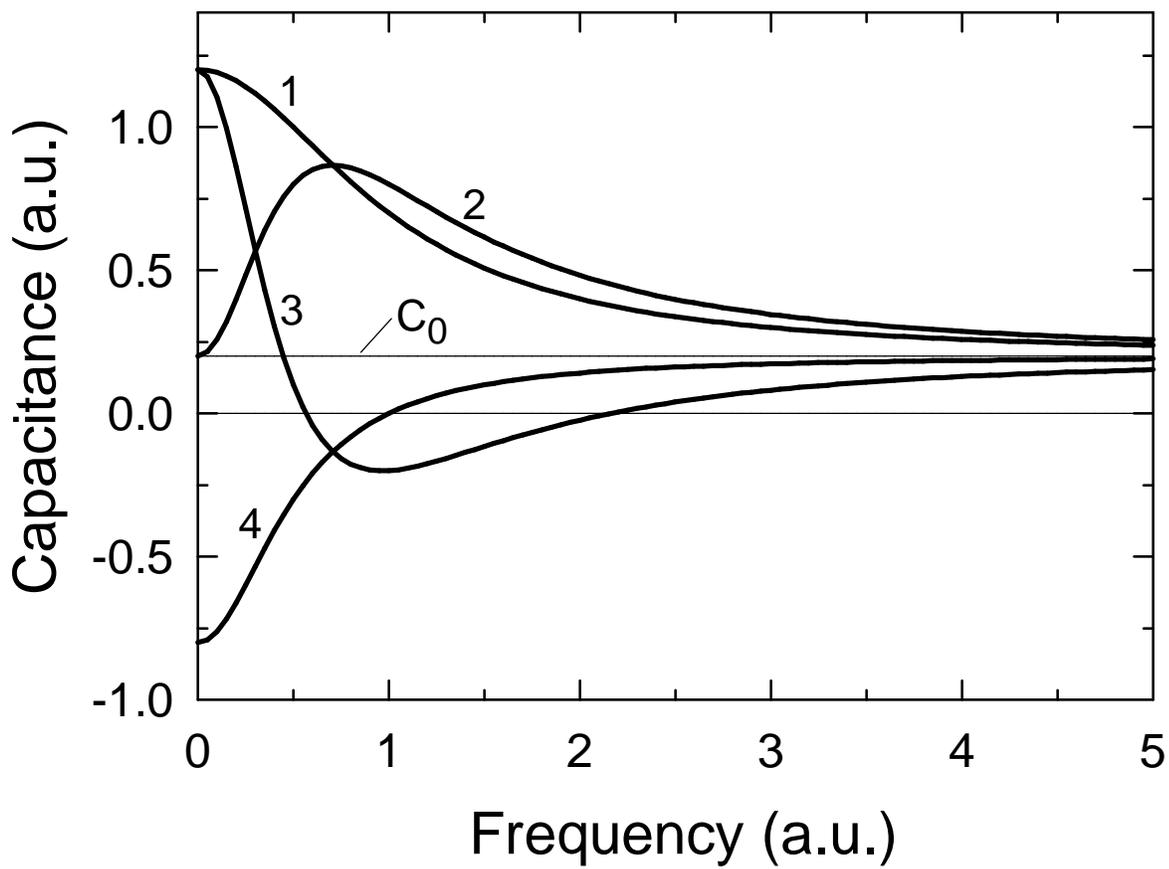
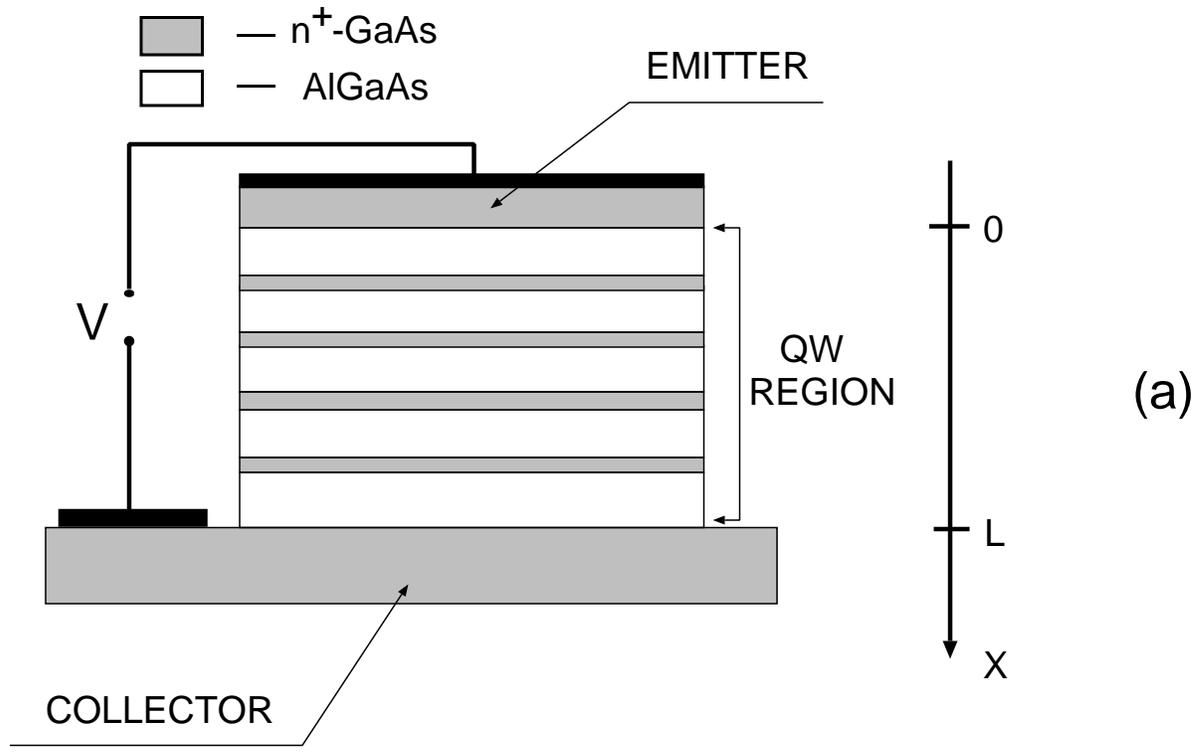
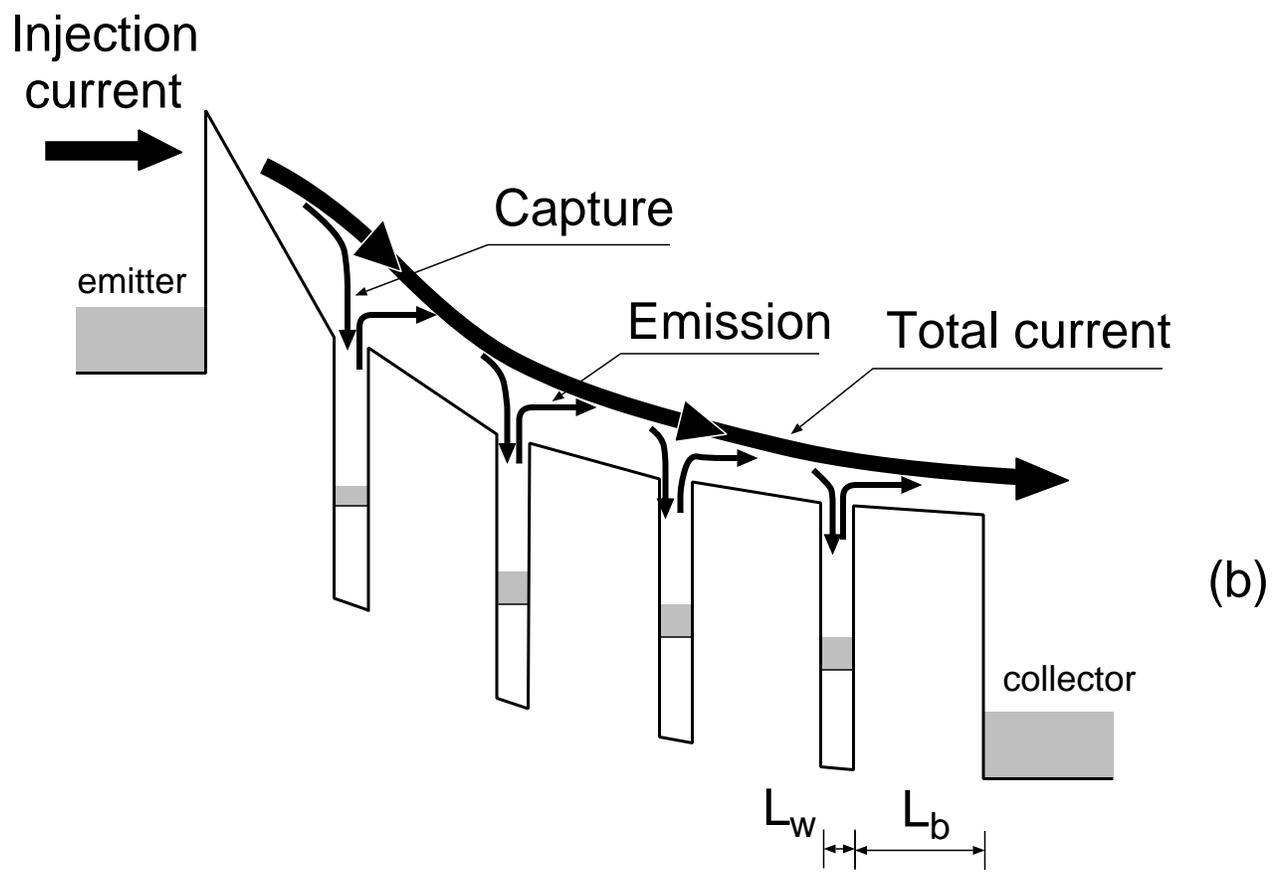
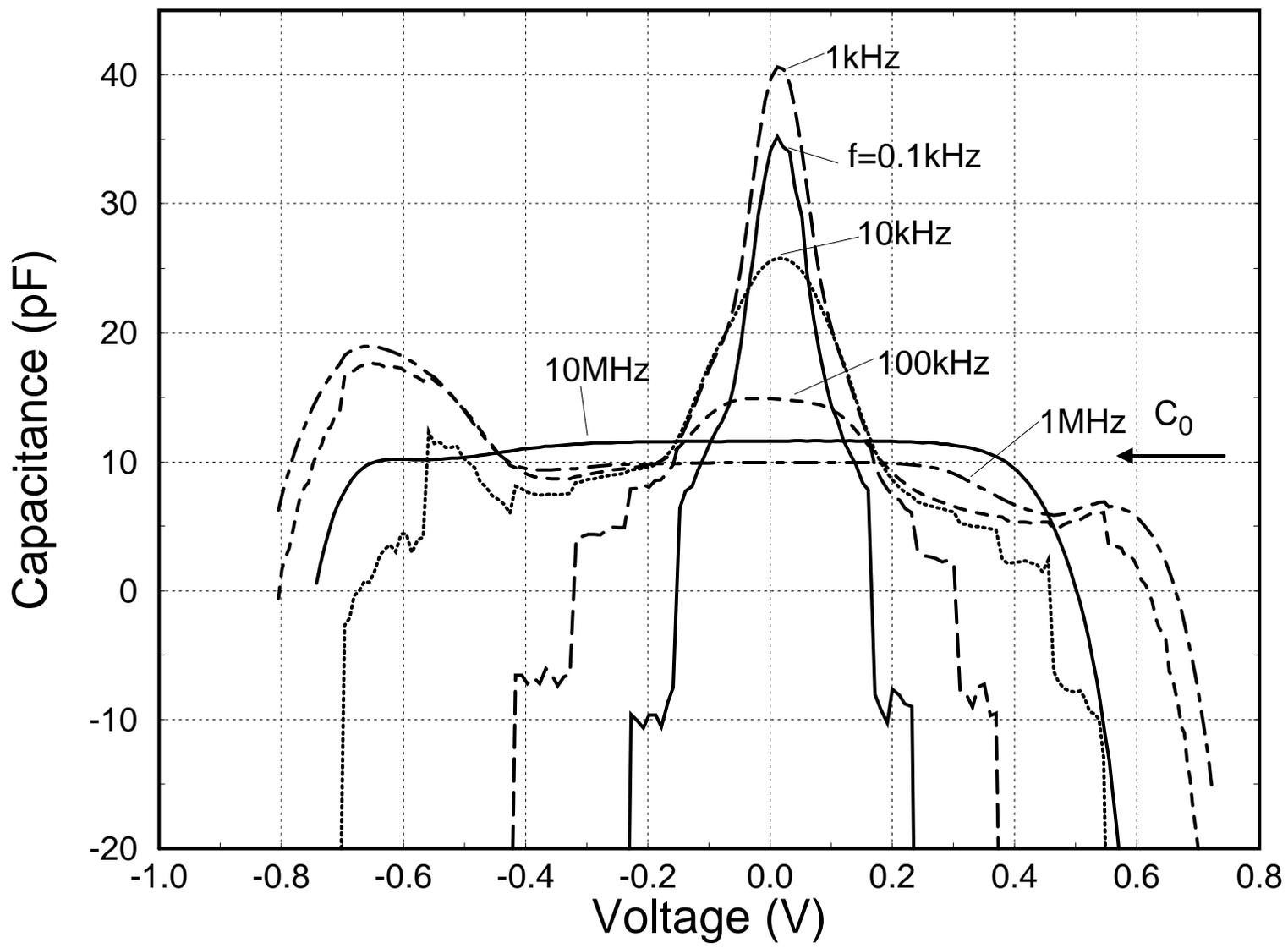
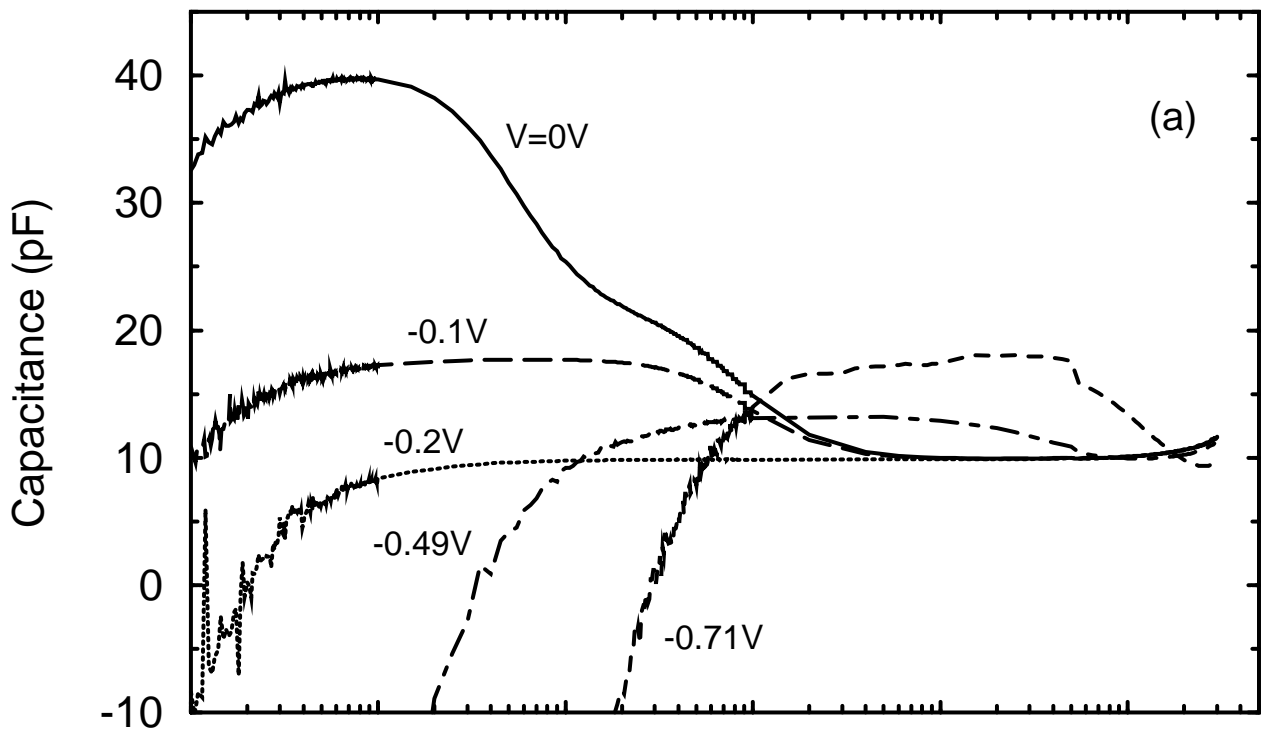
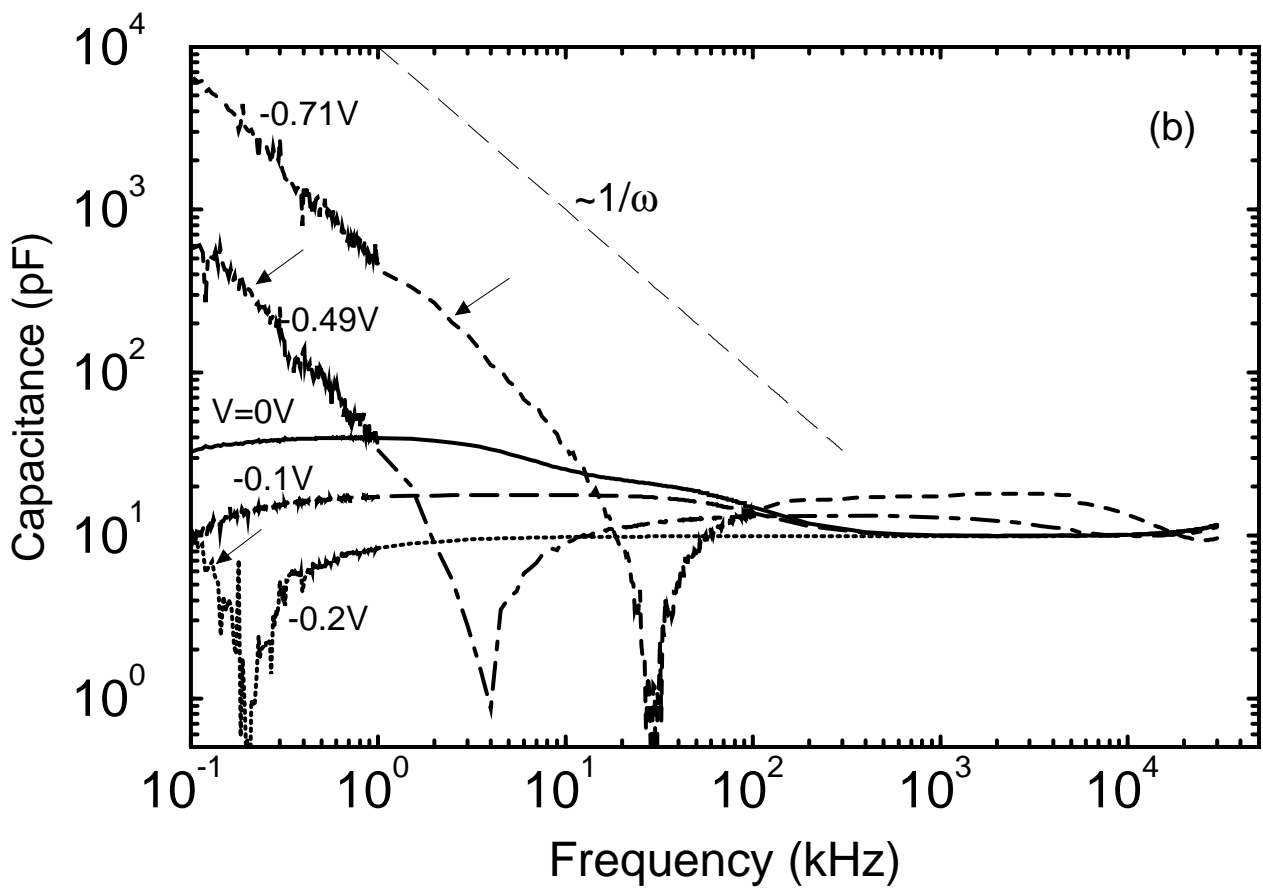
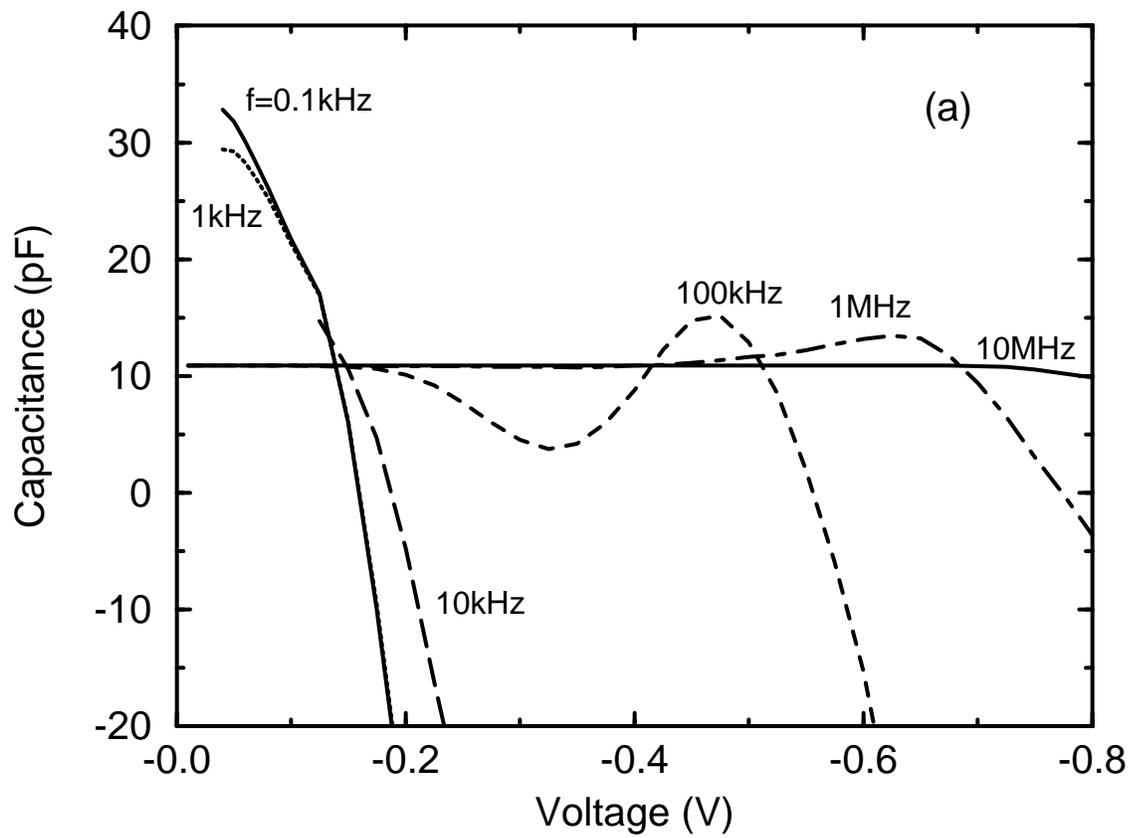
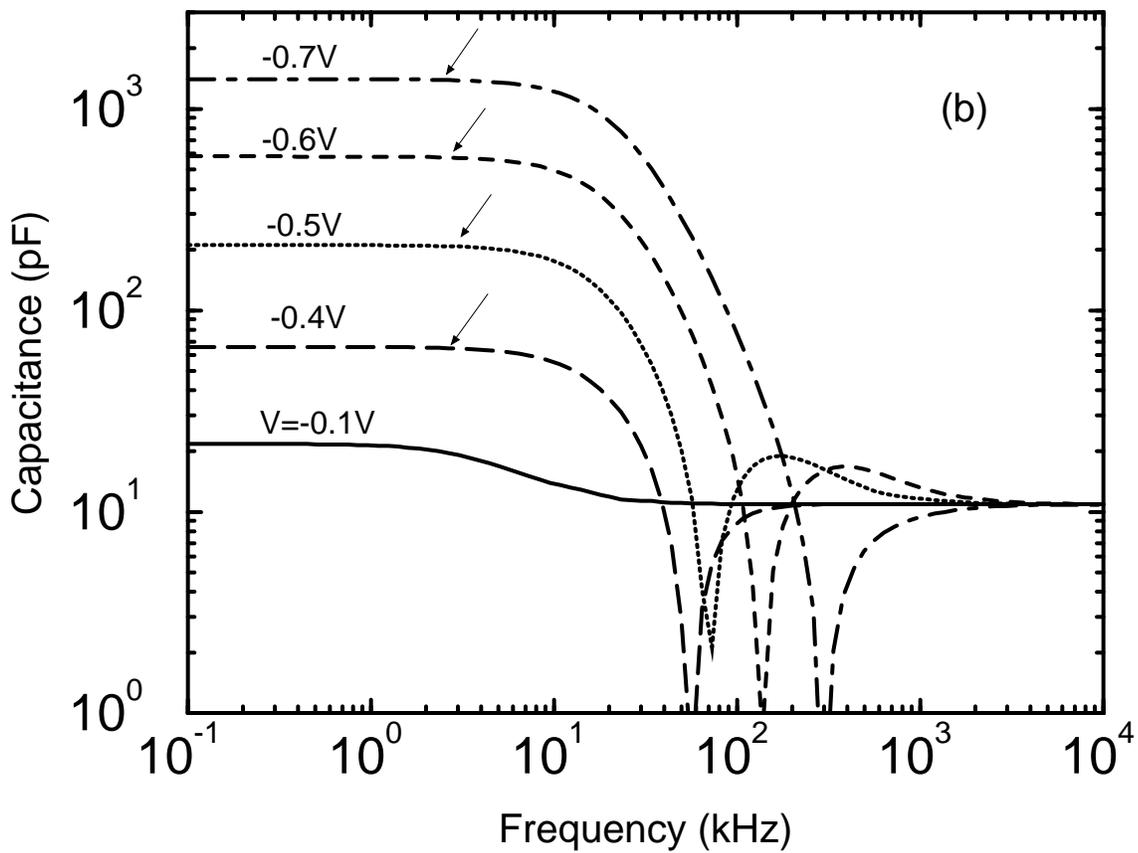
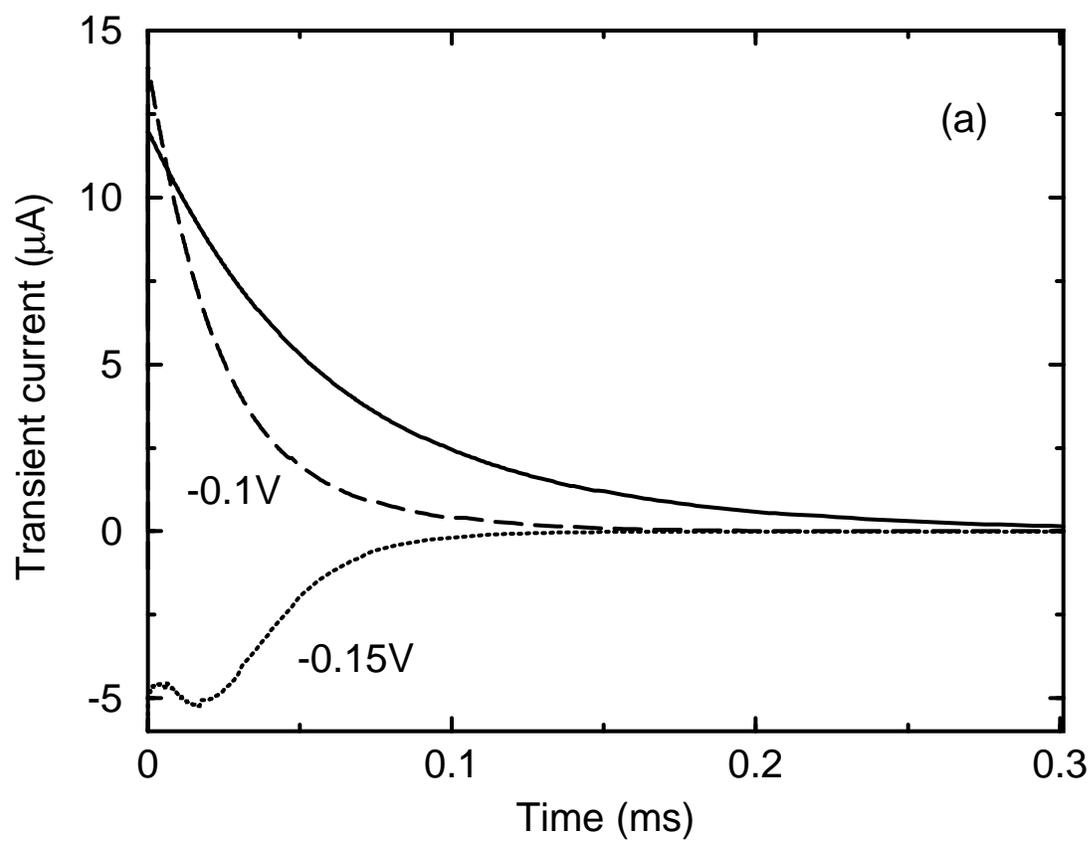
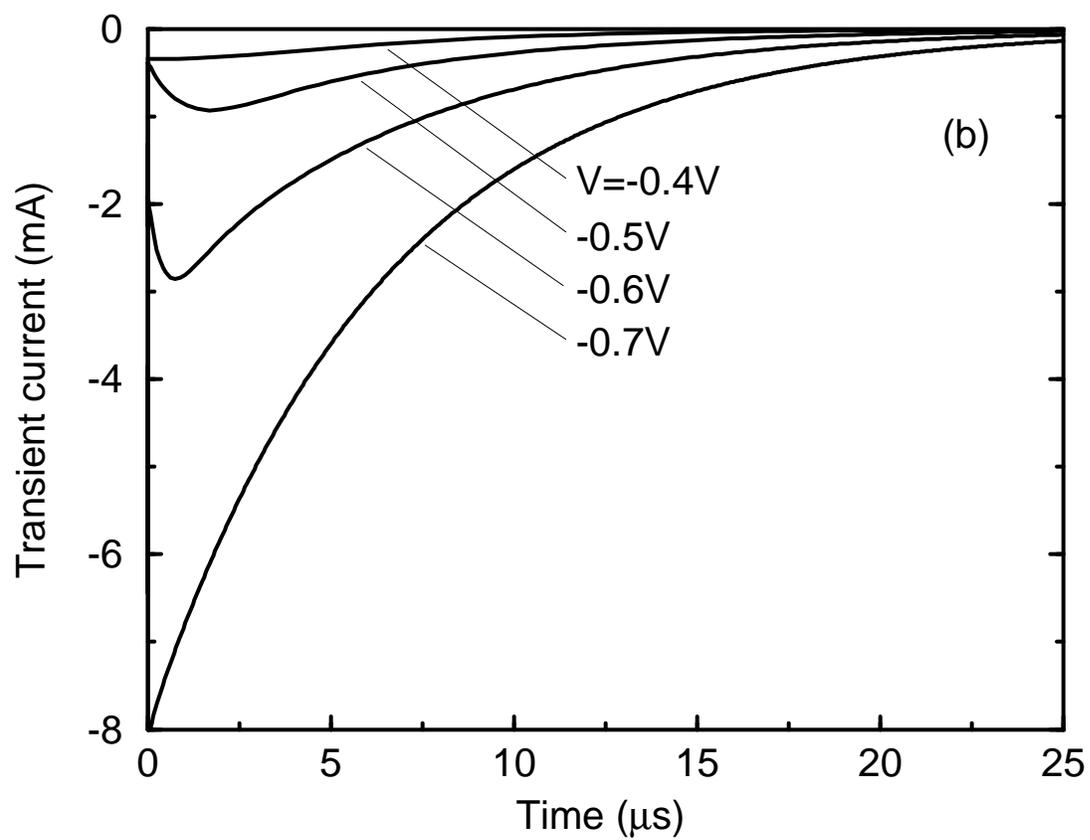
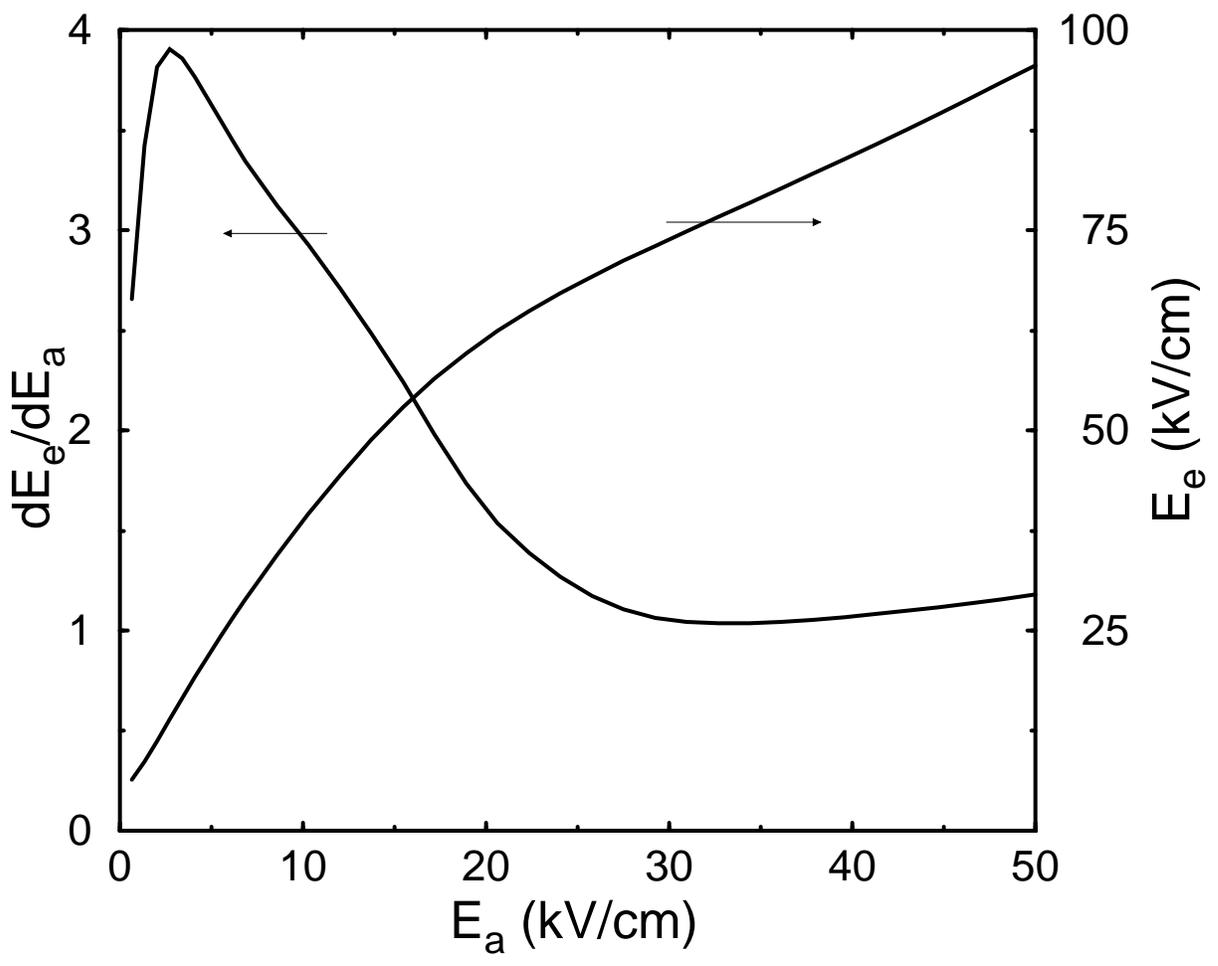
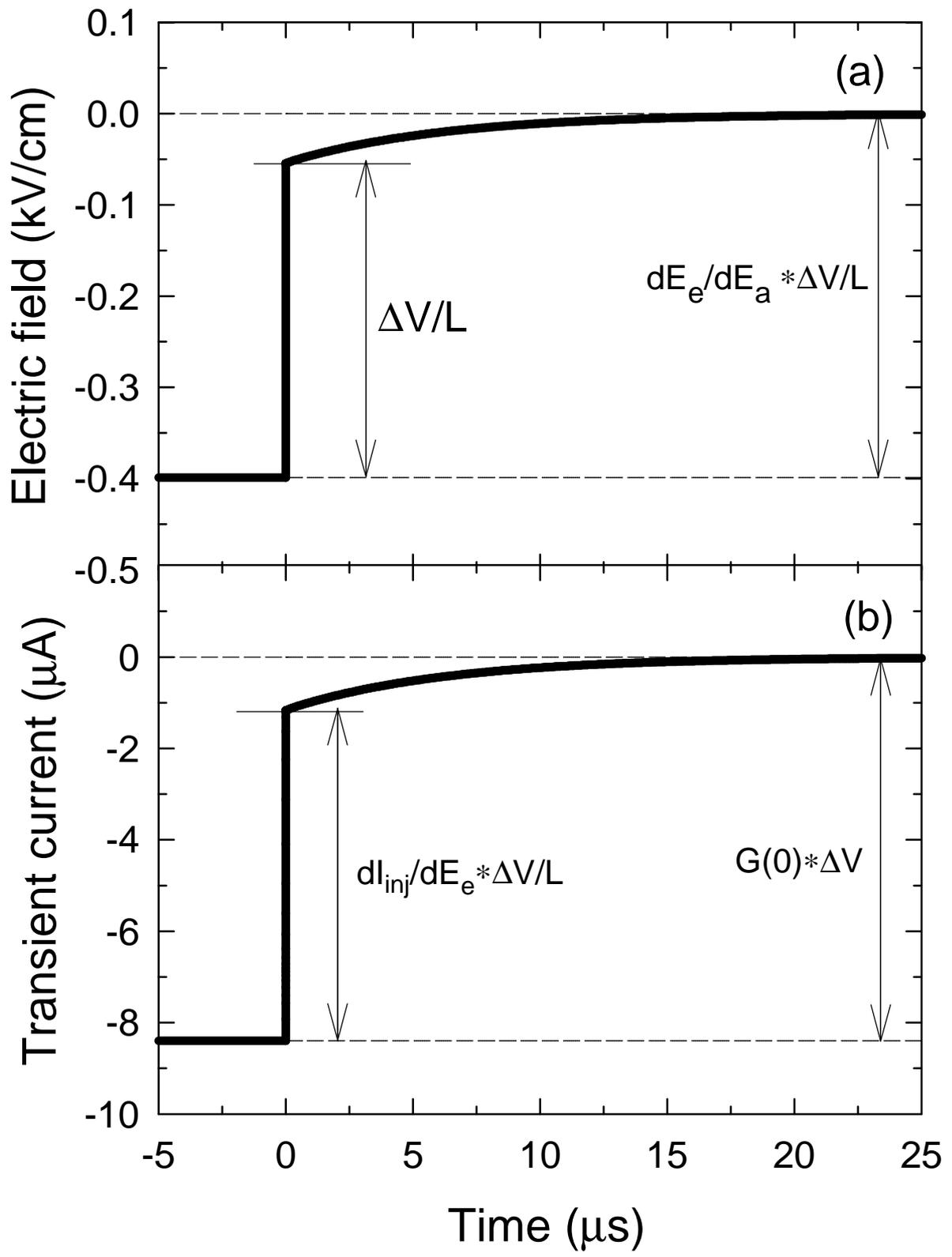
\end{document}